\documentclass[twoside,twocolumn,9pt]{article}
\usepackage{extsizes}
\usepackage[super,sort&compress,comma]{natbib} 
\usepackage[version=3]{mhchem}
\usepackage[left=1.5cm, right=1.5cm, top=1.785cm, bottom=2.0cm]{geometry}
\usepackage{balance}
\usepackage{times,mathptmx}
\usepackage{sectsty}
\usepackage{graphicx} 
\usepackage{lastpage}
\usepackage{amsfonts}
\usepackage[format=plain,justification=justified,singlelinecheck=false,font={stretch=1.125,small,sf},labelfont=bf,labelsep=space]{caption}
\usepackage{float}
\usepackage{fancyhdr}
\usepackage{fnpos}
\usepackage[english]{babel}
\addto{\captionsenglish}{%
  
}
\usepackage{array}
\usepackage{droidsans}
\usepackage{charter}
\usepackage[T1]{fontenc}
\usepackage[usenames,dvipsnames]{xcolor}
\usepackage{setspace}
\usepackage[compact]{titlesec}
\usepackage{hyperref}

\usepackage{epstopdf}

\definecolor{cream}{RGB}{222,217,201}

\begin{document}

\pagestyle{fancy}
\thispagestyle{plain}
\fancypagestyle{plain}{

\fancyhead[C]{\includegraphics[width=18.5cm]{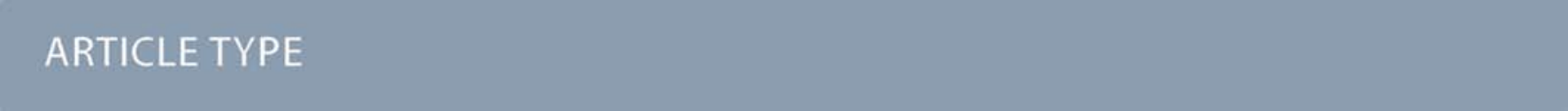}}
\fancyhead[L]{\hspace{0cm}\vspace{1.5cm}\includegraphics[height=30pt]{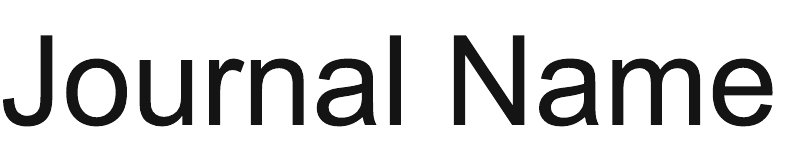}}
\fancyhead[R]{\hspace{0cm}\vspace{1.7cm}\includegraphics[height=55pt]{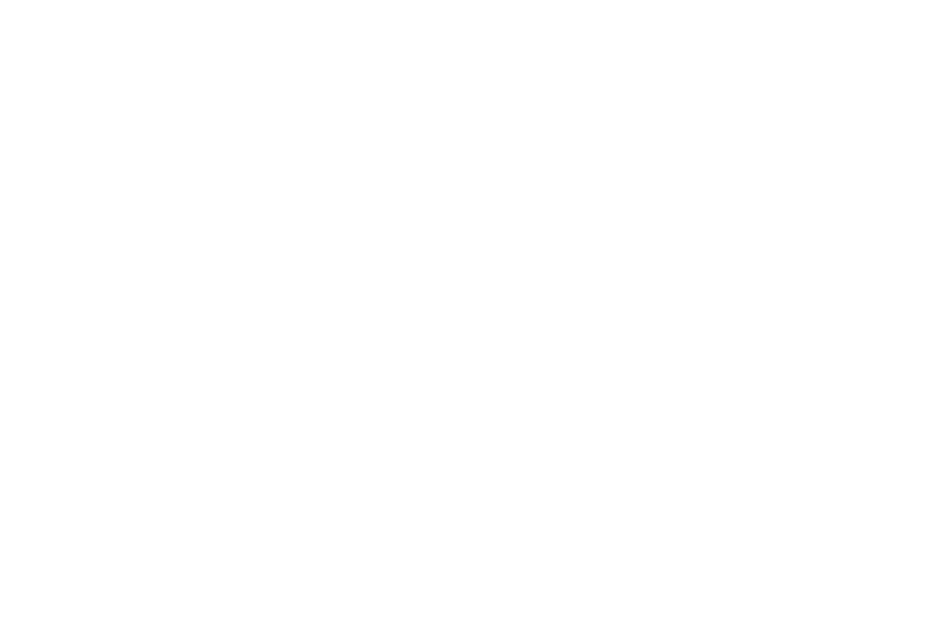}}
\renewcommand{\headrulewidth}{0pt}
}

\makeFNbottom
\makeatletter
\renewcommand\LARGE{\@setfontsize\LARGE{15pt}{17}}
\renewcommand\Large{\@setfontsize\Large{12pt}{14}}
\renewcommand\large{\@setfontsize\large{10pt}{12}}
\renewcommand\footnotesize{\@setfontsize\footnotesize{7pt}{10}}
\makeatother

\renewcommand{\thefootnote}{\fnsymbol{footnote}}
\renewcommand\footnoterule{\vspace*{1pt}%
\color{cream}\hrule width 3.5in height 0.4pt \color{black}\vspace*{5pt}} 
\setcounter{secnumdepth}{5}

\makeatletter 
\renewcommand\@biblabel[1]{#1}            
\renewcommand\@makefntext[1]%
{\noindent\makebox[0pt][r]{\@thefnmark\,}#1}
\makeatother 
\renewcommand{\figurename}{\small{Fig.}~}
\sectionfont{\sffamily\Large}
\subsectionfont{\normalsize}
\subsubsectionfont{\bf}
\setstretch{1.125} 
\setlength{\skip\footins}{0.8cm}
\setlength{\footnotesep}{0.25cm}
\setlength{\jot}{10pt}
\titlespacing*{\section}{0pt}{4pt}{4pt}
\titlespacing*{\subsection}{0pt}{15pt}{1pt}

\newcommand{\be}{\begin{equation}}
\newcommand{\ee}{\end{equation}}
\newcommand{\ba}{\begin{eqnarray}}
\newcommand{\ea}{\end{eqnarray}}
\newcommand{\la}{\langle}
\newcommand{\ra}{\rangle}
\newcommand{\rr}{\mathbf{r}}
\newcommand{\qq}{\mathbf{q}}

\fancyfoot{}
\fancyfoot[LO,RE]{\vspace{-7.1pt}\includegraphics[height=9pt]{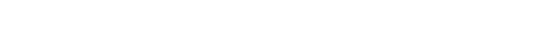}}
\fancyfoot[CO]{\vspace{-7.1pt}\hspace{13.2cm}\includegraphics{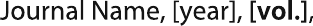}}
\fancyfoot[CE]{\vspace{-7.2pt}\hspace{-14.2cm}\includegraphics{head_foot/RF}}
\fancyfoot[RO]{\footnotesize{\sffamily{1--\pageref{LastPage} ~\textbar  \hspace{2pt}\thepage}}}
\fancyfoot[LE]{\footnotesize{\sffamily{\thepage~\textbar\hspace{3.45cm} 1--\pageref{LastPage}}}}
\fancyhead{}
\renewcommand{\headrulewidth}{0pt} 
\renewcommand{\footrulewidth}{0pt}
\setlength{\arrayrulewidth}{1pt}
\setlength{\columnsep}{6.5mm}
\setlength\bibsep{1pt}

\newcommand{\joe}[1]{\textcolor{blue} {#1}} 
\newcommand{\joec}[1]{\textcolor{blue} {[Joe: #1]}} 
\newcommand*{\cmb}[1]{{\color{BurntOrange}{#1}}}

\makeatletter 
\newlength{\figrulesep} 
\setlength{\figrulesep}{0.5\textfloatsep} 

\newcommand{\topfigrule}{\vspace*{-1pt}%
\noindent{\color{cream}\rule[-\figrulesep]{\columnwidth}{1.5pt}} }

\newcommand{\botfigrule}{\vspace*{-2pt}%
\noindent{\color{cream}\rule[\figrulesep]{\columnwidth}{1.5pt}} }

\newcommand{\dblfigrule}{\vspace*{-1pt}%
\noindent{\color{cream}\rule[-\figrulesep]{\textwidth}{1.5pt}} }

\makeatother

\twocolumn[
  \begin{@twocolumnfalse}
\vspace{3cm}
\sffamily
\begin{tabular}{m{4.5cm} p{13.5cm} }

\includegraphics{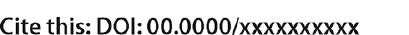} & \noindent\LARGE{\textbf{Active interaction switching controls the dynamic heterogeneity of soft colloidal dispersions}} \\
\vspace{0.3cm} & \vspace{0.3cm} \\

 & \noindent\large{Michael Bley,$^{\textmd{a}}$ Pablo I. Hurtado,$^{\textmd{b}\textmd{c}}$ Joachim Dzubiella,$^{\ast}$$^{\textmd{a}}$ and Arturo Moncho-Jord\'a$^{\ast}$$^{\textmd{c}\textmd{d}}$} \\

\includegraphics{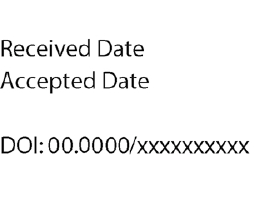} & \noindent\normalsize{

We employ Reactive Dynamical Density Functional Theory (R-DDFT) and Reactive Brownian Dynamics (R-BD) simulations to investigate the dynamics of a suspension of active soft Gaussian colloids with binary interaction switching, i.e., a one-component colloidal system in which every particle stochastically switches at predefined rates between two interaction states with different mobility. Using R-DDFT we extend a theory previously developed to access the dynamics of inhomogeneous liquids [Archer \textit{et al., Phys. Rev. E}, 2007, \textbf{75}, 040501] to study the influence of the switching activity on the self and distinct part of the Van Hove function in bulk solution, and determine the corresponding mean squared displacement of the switching particles. Our results demonstrate that, even though the average diffusion coefficient is not affected by the switching activity, it significantly modifies the non-equilibrium dynamics and diffusion coefficients of the individual particles, leading to a crossover from short to long times, with a regime for intermediate times showing anomalous diffusion. In addition, the self-part of the van Hove function has a Gaussian form at short and long times, but becomes non-Gaussian at intermediates ones, having a crossover between short and large displacements. The corresponding self-intermediate scattering function shows the two-step relaxation patters typically observed in soft materials with heterogeneous dynamics such as glasses and gels. We also introduce a phenomenological Continuous Time Random Walk (CTRW) theory to understand the heterogeneous diffusion of this system. R-DDFT results are in excellent agreement with R-BD simulations and the analytical predictions of CTRW theory, thus confirming that R-DDFT constitutes a powerful method to investigate not only the structure and phase behavior, but also the dynamical properties of non-equilibrium active switching colloidal suspensions.}
\end{tabular}
 \end{@twocolumnfalse} \vspace{0.6cm}

  ]

\renewcommand*\rmdefault{bch}\normalfont\upshape
\rmfamily
\section*{}
\vspace{-1cm}


\footnotetext{\textit{$^{a}$~Physikalisches Institut, Albert-Ludwigs-Universit\"at Freiburg, Hermann-Herder Stra{\ss}e 3, D-79104 Freiburg, Germany.}}
\footnotetext{\textit{$^{b}$~Departamento de Electromagnetismo y F\'{\i}sica de la Materia, Universidad de Granada, Campus Fuentenueva S/N, 18071 Granada, Spain.}}
\footnotetext{\textit{$^{c}$~Institute Carlos I for Theoretical and Computational Physics, Facultad de Ciencias, Universidad de Granada, Campus Fuentenueva S/N, 18071 Granada, Spain.}}
\footnotetext{\textit{$^{d}$~Departamento de F\'{\i}sica Aplicada, Universidad de Granada, Campus Fuentenueva S/N, 18071 Granada, Spain.}}

%
%



\section{Introduction}

Responsive materials like polymers adapt to external stimuli (i.e. temperature, pH, or food) and change their shape and conformation accordingly \cite{Bajpai2008, Stuart2010, Herves2012, Wu2012, Theato2013}. The environment-dependent size of responsive materials can change by a factor of two or three typically \cite{Schroeder2015}. Examples of this are globular proteins, in which their conformation and phase behavior can be tuned by changes in the local environment and protein-protein attractive interaction~\cite{Stegen2015,Stegen2015a}. These colloidal systems and their responsiveness can be used for tailor-made functionality such as core-shell nano-reactors for selective catalysis \cite{Herves2012, Wu2012} or controlled drug release \cite{Bajpai2008, Stuart2010}, but also lay the foundation for adaptive and intelligent systems \cite{Andreas}. Due to the switching of properties such as size or interactions, these compounds can exhibit a complex and rich anomalous diffusion behavior different to the classical Einstein-Smoluchowski picture of a random walk leading to a mean squared displacement (MSD) entirely linear in time $\propto t$ \cite{Einstein1905, Frey2005}. The effective diffusion coefficient $D$ thus changes with the observed time scale. This anomalous diffusion appears for example in various biological systems and during the transport of particles through membranes \cite{Metzler2000, Wang2009, Wang2012, Hoefling2013, Kwon2014, Chubynsky2014, Ghosh2015, Jain2016, Metzler2017, Gnan2019, Yamamoto2021}, where both sub- and superdiffusive regions are observed. 

Heterogeneous diffusion is also reported in a variety of amorphous materials, from low density gels to dense glasses~\cite{Richert2002, Hurtado2007, Chaudhuri2007, Berthier2011, Levis2014, Wang2020}. In most cases such dynamic heterogeneity can be explained by the presence of different particle arrest mechanisms at play. These mechanisms may range from the bonding of particles to the giant (percolating) component of a low density physical gel, which traps particles in a localized region during a long time~\cite{Gado2004,Zaccarelli2005,Gado2007,Zaccarelli2007}, to the steric hindrance induced by crowding effects in glasses, which lead to strong dynamic heterogeneities~\cite{Chaudhuri2007, Berthier2011}. Moreover, different arrest mechanisms can also compete, leading to complex relaxation behaviors~\cite{Zaccarelli2009,Chaudhuri2010,Khalil2014,Chaudhuri2015}. In all cases the observed heterogeneous diffusion is related to the coexistence of fast and slow diffusing particles in the system of interest. These observations suggest that active colloidal systems where particles can stochastically change their internal state, and hence their mobility, may also lead to heterogeneous diffusion properties.


In this work, we investigate a non-equilibrium active system formed by soft colloids in which individual particles stochastically switch between two interaction states (or sizes), denoted as big (b) and small (s), at predefined rates $k_\textmd{bs}$ and $k_\textmd{sb}$. Such a system constitutes a good model for bistable bacteria that use switching to tune structural and dynamical heterogeneities for their function~\cite{Balaban, Dubnau}, as well as for soft active or vesicles fluctuating between two states~\cite{oscillating,Heuser,DNA_hydrogel,breathing}. It could also be applied to study the structure and phase behavior of conformationally fluctuating biopolymers~\cite{Shin,IDP_switching,helical_switching}, in particular two-state proteins switching between native and non-native states~\cite{Stegen2015,Stegen2015a}. In future it may be extendable to even study soft micromachines with a programmable morphology~\cite{Huang2016}.

In our previous works, we studied the structural properties and phase behavior of this active switching system by using a non-equilibrium reactive density functional theory (R-DDFT)~\cite{Glotzer1995, Lutsko2016} and reactive Brownian dynamics simulations (R-BD)~\cite{MonchoJorda2020, Bley2021}. Flavors of R-DDFT have been recently applied to predict the propagation of virus spreading~\cite{Vrugt2020a, Vrugt2021} and for describing the growth of tumors~\cite{AndyPRE2018}. High switching rates lead to mixing of systems which phase separate in equilibrium conditions, whereas low, non-zero switching rates lead to the observation of temporal clusters representing local and temporal phase separation. Here, we extend the R-DDFT framework by means of the test-particle method developed by Archer \textit{et al.}~\cite{Archer2007} to investigate the non-equilibrium steady-state dynamics of actively switching particles in the bulk. Different, but constant diffusion coefficients are assigned to the two particle sizes. This method provides a new pathway for accessing the dynamics of particles switching between two diffusion coefficients. Moreover, we develop a phenomenological Continuous Time Random Walk theory (CTRW)~\cite{Montroll1965, Hurtado2007, Chaudhuri2007} to describe the heterogeneous dynamics of this system. Whereas for R-BD the MSD and the diffusive behavior can be determined directly from the simulated trajectories even at non-equilibrium, accessing the dynamics through R-DDFT requires the calculation of the van-Hove distribution of displacements, $G(\mathbf{r},t)$ \cite{VanHove1954, Hansen2013}. This function, defined as the probability density of finding a particle at time $t$ at location $\mathbf{r}$ from the origin given that there was a particle at the origin at time $t=0$, characterizes dynamical phenomena on a nanoscopic scale. It is especially important in the study of dynamics involved in liquid-crystal, glass-like  and/or sol-gel transitions \cite{Hurtado2007, Chaudhuri2007, Hunter2012}.

The paper is organized as follows. First, we describe the theoretical frameworks used for accessing the van Hove function and the MSDs (R-DDFT, CTRW, and R-BD). In the second part, we discuss the effects that active switching has on the self and distinct parts of the van Hove functions and on the time evolution of the self-intermediate incoherent scattering functions. Finally, the MSDs obtained with R-DDFT are compared with R-BD and CTRW predictions, reporting good agreement between all three approaches for all switching activities investigated.

\section{Theory}

\subsection{Reactive Dynamical Density Functional Theory for active switching colloids.}

We consider an active system in which each individual particle can instantaneously switch between two possible states of different size: big (b) or a small (s). Particles in state $\textmd{b}$ spontaneously convert into state $\textmd{s}$ at some pre-defined rate $k_\textmd{bs}$ (units of time$^{-1}$). Similarly, particles in state $\textmd{s}$ switch into particles of state $\textmd{b}$ at rate $k_\textmd{sb}$. At certain time, the active system resembles a binary mixture of big and small particles. 

The interaction between a pair of Gaussian particles is given by~\cite{Stillinger1976}
\begin{equation}
\label{ugaussian}
\beta u_{ij}=\epsilon_{ij}e^{-r^2/\sigma_{ij}^{2}}  \ \ \ \text{with}\ \ \ i,j=\textmd{s},\textmd{b},
\end{equation}
where $r$ is the interparticle distance, $\beta=1/k_BT$ ($k_B$ is the Boltzmann constant and $T$ the absolute temperature), $\epsilon_{ij}>0$ denotes the strength of the $i$-$j$ pair interactions, and $\sigma_{ij}$ represents the range (we will denote $\sigma_\textmd{bb}$ and $\sigma_\textmd{ss}$ by $\sigma_\textmd{b}$ and $\sigma_\textmd{s}$ respectively, to simplify notation). These soft pair potentials remain finite for any interparticle distance, so particles can interpenetrate each other. The Gaussian pair potential represents a generic model for polymers and soft colloidal hydrogels~\cite{SoftColloid,Louis2000,Scotti} and cells~\cite{Winkler}, and served as a useful and insightful test system in DDFT applications~\cite{Dzubiella2003,Archer2005}.

If the system is immersed inside an external field, the density profiles become inhomogeneous. We denote $u_i^\textmd{ext}(\mathbf{r})$ ($i=\textmd{b},\textmd{s}$) as the external potentials acting on the big and small colloids at position $\mathbf{r}$ (we assume a general case in which the external potential is different for each particle state). These potentials can be caused by applied external forces (such as electrostatic or gravitational fields) or simply represent the effect of confining walls or a single fixed particle. We denote $\rho_\textmd{b}(\mathbf{r},t)$ and $\rho_\textmd{s}(\mathbf{r},t)$ as the number density of colloids in the big and small state at position $\mathbf{r}$ at time $t$, respectively.

The time evolution of a non-equilibrium system of active switching Brownian particles can be predicted by the so-called Reactive Dynamical Density Functional Theory (R-DDFT). Within this theoretical framework, the time evolution of $\rho_i(\mathbf{r},t)$ ($i=\textmd{b},\textmd{s}$) obey the following set of differential equations~\cite{Glotzer1995,Lutsko2016,AndyPRE2018,Liu2020,MonchoJorda2020,Bley2021}
\begin{equation}
\begin{cases}
\frac{\partial \rho_\textmd{b}(\mathbf{r},t)}{\partial t}=-\nabla \cdot \mathbf{J}_\textmd{b} + k_\textmd{sb}\rho_\textmd{s}(\mathbf{r},t) - k_\textmd{bs}\rho_\textmd{b}(\mathbf{r},t) \\
\frac{\partial \rho_\textmd{s}(\mathbf{r},t)}{\partial t}=-\nabla \cdot \mathbf{J}_\textmd{s} + k_\textmd{bs}\rho_\textmd{b}(\mathbf{r},t) - k_\textmd{sb}\rho_\textmd{s}(\mathbf{r},t)
\end{cases},
\label{RDDFT}
\end{equation}

The first term on the right side, $-\nabla \cdot\mathbf{J}_i$, provides the change of particle concentrations due to diffusion, where $\mathbf{J}_i$ are the time and position-dependent diffusive fluxes caused by gradient in particle concentrations and chemical potential. They are given by
\begin{equation}
\label{DDFT2}
\mathbf{J}_i=-D_i \left[ \nabla \rho_i(\mathbf{r},t)+\rho_i(\mathbf{r},t)\beta\nabla \big( u_i^\textmd{ext}(\mathbf{r})+ \mu_i^\textmd{ex}(\mathbf{r},t)\big) \right] \ \ \ i=\textmd{b}, \textmd{s},
\end{equation}
where $D_i$ are the diffusion constant of component $i$ (which is assumed to be independent on the specific location of the particles), and $\mu_i^\textmd{ex}(\mathbf{r},t)=\frac{\delta F^\textmd{ex}[\{ \rho_i(\mathbf{r},t) \}]}{\delta \rho_i(\mathbf{r},t)}$. $F^\textmd{ex}[\{ \rho_i(\mathbf{r},t)]$ is the equilibrium excess free energy functional with the equilibrium density profiles replaced by the non-equilibrium ones $\rho_i(\mathbf{r},t)$. The equilibrium and non-equilibrium properties of soft Gaussian particles described by Eq.~\ref{ugaussian} are well represented by the mean-field excess free energy functional for colloidal mixtures of two states,
\begin{equation}
\label{freeenergy}
F^\textmd{ex}[\{\rho_i(\mathbf{r}) \}]=\frac{1}{2}\sum_{i,j=\textmd{b},\textmd{s}}\iint \rho_i(\mathbf{r}) \rho_j(\mathbf{r}^{\prime}) u_{ij}(|\mathbf{r}-\mathbf{r}^{\prime}|)d\mathbf{r}d\mathbf{r}^{\prime}.
\end{equation}

The other two terms on the right hand of Eq.~\ref{RDDFT} account for the production and disappearance of each particle state due to active switching. This process occurs locally, so the conversion rate of colloids in the big state into the small state at some specific location $\mathbf{r}$ only depends on the local concentrations of both species at position $\mathbf{r}$.

In the region far away from the external perturbation (bulk), the density profiles tend to be homogeneous. We define the composition parameters of the mixture as
\begin{equation}
x_\textmd{s}(t)=\frac{\rho_\textmd{s}^{bulk}(t)}{\rho_\textmd{b}^{bulk}(t)+\rho_\textmd{s}^{bulk}(t)} \ \ \ , \ \ \ x_\textmd{b}(t)=1-x_\textmd{s}(t).
\end{equation}
In this work, we select the composition at time $t=0$ such that
\begin{equation}
\label{condition}
\frac{k_\textmd{bs}}{k_\textmd{sb}}=\frac{x_\textmd{s}(0)}{x_\textmd{b}(0)}.
\end{equation}
This condition implies that the relative composition of the mixture of states in the bulk solution remains constant for all times, even though the inhomogeneous properties of the system exposed to external potentials are still affected by the non-equilibrium switching.

The switching activity, $a$, is defined as the ratio between the typical characteristic big-to-small conversion time, $\tau_{switch}=k_\textmd{bs}^{-1}$, and the Brownian diffusion time for small particles, $\tau_B=\sigma_\textmd{s}^2/D_\textmd{s}$. Therefore,
\begin{equation}
a=\frac{\tau_B}{\tau_{switch}}=\frac{k_\textmd{bs}\sigma_\textmd{s}^2}{D_\textmd{s}}.
\end{equation}

In the absence of switching activity ($a=0$), the R-DDFT equations reduce to the classical DDFT equations for non-active binary mixture of Brownian colloids, i.e. $\partial \rho_i(\mathbf{r},t)/\partial t=-\nabla \cdot \mathbf{J}_i$ ($i=\textmd{b},\textmd{s}$)~\cite{Marconi1999,Archer2004,Vrugt2020}. For $a \ll 1$, the \ce{b <=> s} conversion rate is very slow, so the time evolution of the density profiles is dominated by the diffusion. In this case, switching events happening at some specific location are scarce, and the corresponding change in particle concentrations is rapidly compensated by the diffusive fluxes that balance the effect of the activity. In the opposite limit, $a \gg 1$, the switching rate is so large that the diffusion is not fast enough to compensate its effects, so the exchange activity dominates. In this limit, particles in states $\textmd{b}$ and $\textmd{s}$ cannot be distinguished because they do not have enough time to diffuse and reorganize according to the applied external potentials. Consequently, both density profiles converge to each other, and the nonequilibrium system behaves as an effective one-component system that can be described by a single effective pair potential in equilibrium~\cite{MonchoJorda2020,Bley2021}.

If the applied external potentials do not depend on time, the R-DDFT equations lead in the limit $t\rightarrow \infty$ to steady-state density profiles, $\lim_{t\rightarrow \infty}\rho_i(\mathbf{r},t)=\rho_i(\mathbf{r})$. For $a=0$, this final steady state corresponds to the equilibrium, which means that the resulting density profiles are the ones of an equilibrium binary mixture. However, it is important to emphasize that this is not the case for $a>0$. For active systems, the final steady-state density profiles are not the equilibrium ones, even though they are time-independent. It may be shown that thermodynamic properties depend on the diffusion coefficients of the particles, reflecting the fact that the system is not in equilibrium~\cite{MonchoJorda2020}.

The microstructure of the system in bulk suspension is also affected by the switching activity. The non-equilibrium steady-state partial radial distribution functions $g_{ij}(r)$ of the active system can be deduced making use of the Percus test particle route and extending the above described 2-states R-DDFT procedure to a 4-states R-DDFT~\cite{MonchoJorda2020,Bley2021}.

In the next subsections we describe how the R-DDFT method can be generalized to access the dynamics of active suspensions of switching Gaussian colloids. For this purpose, we start describing the simpler non-active one-component system and then extend the procedure to incorporate the active switching.

\subsection{Dynamics of a non-active one-component system of Gaussian colloids}

Here, we consider a homogeneous one-component system formed by $N$ particles inside a volume $V$ at temperature $T$. The bulk number density is $\rho^{bulk}=N/V$. Particles interact via a pairwise interaction potential $u(r)$. The van Hove distribution of displacements for a one-component uniform fluid is defined as~\cite{VanHove1954,Hopkins2010,Hansen2013}
\begin{equation}
\label{Gtotal}
G(\mathbf{r},t)=\frac{1}{N}\biggl< \sum_{\mu=1}^N\sum_{\nu=1}^N \delta(\mathbf{r}-\mathbf{r}_{\nu}(t)+\mathbf{r}_{\mu}(0))\biggr>.
\end{equation}
where indexes $\mu$ and $\nu$ run over the $N$ particles of the system, and $\langle \cdots \rangle$ represents the ensemble average. The physical interpretation of the van Hove function is that $G(\mathbf{r},t)d\mathbf{r}$ is the number of particles within a volume $d\mathbf{r}$ around a point $\mathbf{r}$ at time $t$ given that there was a particle located at the origin at time $t = 0$. It splits into \textit{self} and \textit{distinct} parts that correspond to the possibilities that $\mu$ and $\nu$ are the same particle or different ones, $G(\mathbf{r},t)=G^{self}(\mathbf{r},t)+G^{dist}(\mathbf{r},t)$, where
\begin{eqnarray}
\label{Gselfdef}
G^{self}(\mathbf{r},t)&=&\frac{1}{N}\biggl< \sum_{\mu=1}^N \delta(\mathbf{r}-\mathbf{r}_{\mu}(t)+\mathbf{r}_{\mu}(0))\biggr> \\
G^{dist}(\mathbf{r},t)&=&\frac{1}{N}\biggl< \sum_{\mu=1}^N\sum_{\nu \neq \mu}^N \delta(\mathbf{r}-\mathbf{r}_{\nu}(t)+\mathbf{r}_{\mu}(0))\biggr>.
\end{eqnarray}

For $t=0$, we find that the self-part is
\begin{equation}
G^{self}(\mathbf{r},0)=\frac{1}{N}\biggl< \sum_{\mu=1}^N \delta(\mathbf{r})\biggr>=\frac{1}{N}N\delta(\mathbf{r})\langle 1 \rangle = \delta(\mathbf{r}),
\end{equation}
which means that the test particle is located at the origin $\mathbf{r}=0$ at time $t=0$. The distinct-part at $t=0$ is connected to the equilibrium 2-particle density, $\rho_N^{(2)}(\mathbf{r},\mathbf{r^{\prime}})$, by~\cite{Hansen2013}
\begin{eqnarray}
G^{dist}(\mathbf{r},0)=\frac{1}{N}\biggl< \sum_{\mu=1}^N\sum_{\nu \neq \mu}^N\int \delta(\mathbf{r}+\mathbf{r}^{\prime}-\mathbf{r}_{\nu}(0))\delta(\mathbf{r}^{\prime}-\mathbf{r}_{\mu}(0))d\mathbf{r}^{\prime}\biggr>= \nonumber \\
=\frac{1}{N}\int\rho_N^{(2)}(\mathbf{r}+\mathbf{r}^{\prime},\mathbf{r}^{\prime})d\mathbf{r}^{\prime}=\frac{1}{N}(\rho^{bulk})^2\int g(r)d\mathbf{r}^{\prime}=\rho^{bulk}g(r)
\end{eqnarray}
where we assumed that the system is homogeneous and isotropic, and $g(r)$ is the equilibrium radial distribution function. 

The standard Density Functional Theory (DFT) together with the Percus' test particle route can be used to access $g(r)$~\cite{Percus1962,Hansen2013}: a single test particle is fixed at the origin $r=0$, acting as an external potential for the rest of particles, so $u^{ext}(r)=u(r)$. Solving the DFT equations for the colloidal fluid exposed to the influence of this external potential leads to a inhomogeneous one-body density distribution of colloids around the central one, $\rho(r)$. The corresponding radial distribution function is given by $g(r)=\rho(r)/\rho^{bulk}$.

The van Hove function in bulk for an homogeneous system can be obtained using the DDFT framework, as proposed by Archer {\it et al.}~\cite{Archer2007}. According to their scheme, the system of $N$ particles is separated into two groups that will be called \textit{self} (group 1) and \textit{distinct} (group 2). On the one hand, \textit{self} group consist of only one single test particle located at $r=0$ at time $t=0$. On the other hand, the \textit{distinct} group is formed by the remaining $N-1$ particles around the test particle. With this strategy, our originally one-component system becomes a binary two-component mixture, in which the pair interactions are given by
\begin{equation}
u^{12}(r)=u^{22}(r)=u(r)   \ \ , \ \ \ u^{11}(r)=0.
\end{equation}
Please note that $u^{11}(r)=0$ because a single particle can not interact with itself. This is equivalent to modeling a one-component system, but treating one particle separately from the rest. At time $t=0$, the central particle is located at the origin, which means that the number density of the \textit{self} component is $\rho^1(r,t=0)=\delta(r)$. Conversely, the other $N-1$ particles that form the \textit{distinct} component are initially distributed following the equilibrium distribution, so $\rho^2(r)=\rho^{bulk}g(r)$, where $g(r)$ has been previously determined using DFT within the Test Particle Route~\cite{Percus1962,Hopkins2011,Hansen2013}.

The time evolution of both distributions can be obtained applying classical DDFT to this mixture
\begin{equation}
\label{DDFTi}
\frac{\partial \rho^\alpha}{\partial t}=-\nabla \cdot \mathbf{J}^\alpha \ \ \ \ \ \alpha=1, 2,
\end{equation}
with
\begin{equation}
\mathbf{J}^\alpha=-D\left[ \nabla \rho^\alpha+\rho^\alpha\beta\nabla \Big(\frac{\delta F^\textmd{ex}}{\delta \rho^\alpha} \Big)\right] \ \ \ \ \ \alpha=1, 2,
\end{equation}
where $D$ is the diffusion coefficient of the particles. 

Solving Eqs.~\ref{DDFTi} provides the time evolution of both particle densities, namely $\rho^1(r,t)$ and $\rho^2(r,t)$. Once this solution has been determined, we can deduce the Van Hove distribution of displacements by splitting it into the self and distinct parts, and identifying each part with the density corresponding profile,
\begin{equation}
G^{self}(r,t)=\rho^1(r,t) \ \ \ , \ \ \ G^{dist}(r,t)=\rho^2(r,t).
\end{equation}

As time increases, $G^{self}(r,t)$ broadens into a Gaussian-shaped curve. Conversely, $G^{dist}(r,t)$ becomes flatter as time evolves. In the limit $t\rightarrow \infty$ or $r\rightarrow \infty$, the self part tends to zero whereas the distinct part converges to the uniform distribution,
\begin{eqnarray}
\lim_{r\rightarrow \infty}G^{self}(r,t)=\lim_{t\rightarrow \infty }G^{self}(r,t)=0, \nonumber \\
\lim_{r\rightarrow \infty }G^{dist}(r,t)=\lim_{t\rightarrow \infty }G^{dist}(r,t)=\rho^{bulk}.
\end{eqnarray}

The MSD of the particles is obtained as an integral of $G^{self}(r,t)$,
\begin{equation}
\label{MSD}
\langle \Delta r^2(t) \rangle = \int r^2G^{self}(r,t)d\mathbf{r}=4\pi \int_0^{\infty}r^4G^{self}(r,t)dr.
\end{equation}

\subsection{Dynamics of an active system of switching Gaussian colloids}

Now, each individual particle is able to switch from two states, $\textmd{b}$ and $\textmd{s}$. The method shown before can be entirely adapted to this new situation. We assume that at time $t=0$ the system has already reached the steady state. The steady-state radial distribution functions of the active system can be obtained using the 4-states R-DDFT with the Test Particle Route~\cite{MonchoJorda2020,Bley2021,Hopkins2011}. This leads to the (non-equilibrium) steady-state radial distribution functions, $g_{ij}(r)$ ($i=\textmd{b},\textmd{s}$), which provide the probability density of finding a particle in state $j$ at a distance $r$ from another particle in the state $i$. Therefore, $\rho_j^{bulk}g_{ij}(r)$ represents the density of particles in state $j$ at a distance $r$ from a central particle in state $i$.

Using a similar procedure than the one followed for the one-component system, we split the system into the self and distinct part (denoted again by superindex 1 and 2). The self-part represents a single test particle located at $r=0$ at time $t=0$, whereas the distinct one corresponds to the rest of particles. Pair interactions between self and distinct particles in states $\textmd{b}$ and $\textmd{s}$ are
\begin{eqnarray}
u^{22}_\textmd{bb}(r)=u^{12}_\textmd{bb}(r)=\epsilon_\textmd{bb}e^{-r^2/\sigma_\textmd{b}^2} \ \ \ \ \ \ \ \ \ \ u^{11}_\textmd{bb}(r)=0 \nonumber \\[-3pt]
u^{22}_\textmd{bs}(r)=u^{12}_\textmd{bs}(r)=\epsilon_\textmd{bs}e^{-r^2/\sigma_\textmd{bs}^2} \ \ \ \ \ \ \ \ \ \ u^{11}_\textmd{bs}(r)=0  \\[-3pt]
u^{22}_\textmd{ss}(r)=u^{12}_\textmd{ss}(r)=\epsilon_\textmd{ss}e^{-r^2/\sigma_\textmd{s}^2} \ \ \ \ \ \ \ \ \ \ u^{11}_\textmd{ss}(r)=0 \nonumber
\end{eqnarray}

If the test particle located at $r=0$ at time $t=0$ is in the $\textmd{b}$-state, then
\begin{equation}
\begin{cases}
\rho^1_\textmd{b}(r,t=0)=\delta(\mathbf{r}) \ \ \ \ , \ \ \ \ \rho^1_\textmd{s}(r,t=0)=0 \\
\rho^2_\textmd{b}(r,t=0)=\rho^{bulk}_\textmd{b}g_\textmd{bb}(r) \ \ \  ,  \ \ \ \rho^2_\textmd{s}(r,t=0)=\rho^{bulk}_\textmd{s}g_\textmd{bs}(r)
\end{cases}
\label{initial1}
\end{equation}

Conversely, if the test central particle is in the $\textmd{s}$-state, the initial conditions are
\begin{equation}
\begin{cases}
\rho^1_\textmd{b}(r,t=0)=0 \ \ \ \ , \ \ \ \ \rho^1_\textmd{s}(r,t=0)=\delta(\mathbf{r}) \\
\rho^2_\textmd{b}(r,t=0)=\rho^{bulk}_\textmd{b}g_\textmd{bs}(r) \ \ \  ,  \ \ \ \rho^2_\textmd{s}(r,t=0)=\rho^{bulk}_\textmd{s}g_\textmd{ss}(r)
\end{cases}
\label{initial2}
\end{equation}


As $t$ increases, the four density profiles evolve in time. Their time evolution is governed by two processes: the diffusion due to gradients of the chemical potential, and the switching events that cause the appearance/disappearance of particle states. The four coupled R-DDFT equations that control this time evolution can be obtained extending Eq.~\ref{RDDFT}
\begin{eqnarray}
\begin{cases}
\frac{\partial \rho^1_\textmd{b}(\mathbf{r},t)}{\partial t}=-\nabla \cdot \mathbf{J}^1_\textmd{b} + k_\textmd{sb}\rho^1_\textmd{s}(\mathbf{r},t) - k_\textmd{bs}\rho^1_\textmd{b}(\mathbf{r},t) \\[3pt]
\frac{\partial \rho^1_\textmd{s}(\mathbf{r},t)}{\partial t}=-\nabla \cdot \mathbf{J}^1_\textmd{s} + k_\textmd{bs}\rho^1_\textmd{b}(\mathbf{r},t) - k_\textmd{sb}\rho^1_\textmd{s}(\mathbf{r},t) \\[3pt]
\frac{\partial \rho^2_\textmd{b}(\mathbf{r},t)}{\partial t}=-\nabla \cdot \mathbf{J}^2_\textmd{b} + k_\textmd{sb}\rho^2_\textmd{s}(\mathbf{r},t) - k_\textmd{bs}\rho^2_\textmd{b}(\mathbf{r},t) \\[3pt]
\frac{\partial \rho^2_\textmd{s}(\mathbf{r},t)}{\partial t}=-\nabla \cdot \mathbf{J}^2_\textmd{s} + k_\textmd{bs}\rho^2_\textmd{b}(\mathbf{r},t) - k_\textmd{sb}\rho^2_\textmd{s}(\mathbf{r},t).
\label{vanHove2}
\end{cases}
\end{eqnarray}

The diffusive fluxes are given by
\begin{equation}
\mathbf{J}^{\alpha}_i=-D_i \left[ \nabla \rho^{\alpha}_i+\rho^{\alpha}_i\beta\nabla  \mu_i^{\textmd{ex},\alpha} \right] \ \ \ \alpha=1,2 \ \ ,\ \ i=\textmd{b},\textmd{s},
\end{equation}
where we have used that the external potentials are zero for a bulk system. The excess chemical potential is obtained from the functional derivative of the total excess free energy, $\mu^{\textmd{ex},\alpha}_i=\delta F^\textmd{ex}/\delta \rho^{\alpha}_i$. For the excess free energy, we make use again of the mean-field approach, useful for Gaussian colloids
\begin{equation}
\label{freeenergy2}
F^\textmd{ex}=\frac{1}{2}\sum_{\alpha,\beta=1,2}\ \sum_{i,j=\textmd{b},\textmd{s}}\iint \rho^\alpha_i(\mathbf{r}) \rho^\beta_j(\mathbf{r}^{\prime}) u^{\alpha\beta}_{ij}(|\mathbf{r}-\mathbf{r}^{\prime}|)d\mathbf{r}d\mathbf{r}^{\prime}.
\end{equation}

Performing the functional differentiation leads to the following explicit expression for the excess chemical potential
\begin{equation}
\mu^{\textmd{ex},\alpha}_i(\mathbf{r},t)=\sum_{\beta=1,2}\sum_{j=\textmd{b},\textmd{s}}\int \rho^{\beta}_j(\mathbf{r}^{\prime},t)u^{\alpha\beta}_{ij}(|\mathbf{r}-\mathbf{r}^{\prime}|)d\mathbf{r}^{\prime}.
\end{equation}

The R-DDFT equations are solved with the initial conditions given by Eq.~\ref{initial1} (if the test central particle in the $\textmd{b}$-state) or Eq.~\ref{initial2} (if the test central particle in the $\textmd{s}$-state). In addition, we need to impose the boundary conditions in $r=0$ and $r\rightarrow \infty$,
\begin{equation}
\begin{cases}
\mathbf{J}^{1}_\textmd{b}(r=0,t)=\mathbf{J}^{1}_\textmd{s}(r=0,t)=\mathbf{J}^{2}_\textmd{b}(r=0,t)=\mathbf{J}^{2}_\textmd{s}(r=0,t)=0 \nonumber \\
\rho^1_\textmd{b}(r\rightarrow \infty,t)=\rho^1_\textmd{s}(r\rightarrow \infty,t)=0  \\
\rho^2_\textmd{b}(r\rightarrow \infty,t)=\rho^{bulk}_\textmd{b} \ \ \ , \ \ \ \rho^2_\textmd{s}(r\rightarrow \infty,t)=\rho^{bulk}_\textmd{s}.  \nonumber
\end{cases}
\label{initial}
\end{equation}

To study the dynamics of this active switching system (Eq.~\ref{Gtotal}), we decompose the self-part of the van Hove function as:
\begin{eqnarray}
G^{self}(\mathbf{r},t)&=&\frac{N_\textmd{b}}{N}\frac{1}{N_\textmd{b}}\biggl< \sum_{\mu=1}^{N_\textmd{b}} \delta(\mathbf{r}-\mathbf{r}_{\mu}(t)+\mathbf{r}_{\mu}(0))\biggr> \nonumber \\
&+& \frac{N_\textmd{s}}{N}\frac{1}{N_\textmd{s}}\biggl< \sum_{\mu=1}^{N_\textmd{s}} \delta(\mathbf{r}-\mathbf{r}_{\mu}(t)+\mathbf{r}_{\mu}(0))\biggr>,
\end{eqnarray}
where $N_\textmd{b}$ and $N_\textmd{s}$ are the total number of particles in $\textmd{b}$ and $\textmd{s}$ states in the steady-state, respectively ($N=N_\textmd{b}$+$N_\textmd{s}$). We define the partial self-part of the van Hove function as~\cite{Kob1995}
\begin{equation}
G^{self}_i(\mathbf{r},t)=\frac{1}{N_i}\biggl< \sum_{\mu=1}^{N_i} \delta(\mathbf{r}-\mathbf{r}_{\mu}(t)+\mathbf{r}_{\mu}(0))\biggr>,
\end{equation}
which is calculated assuming that at time $t=0$ there was a particle in the $i$-state located at $r=0$. Using that $x_{i}=N_i/N$ we find
\begin{equation}
G^{self}(r,t)=x_\textmd{b}G^{self}_\textmd{b}(r,t)+x_\textmd{s}G^{self}_\textmd{s}(r,t).
\end{equation}

We can identify the self-part of the van Hove with the particle densities of component $1$, i.e.
\begin{equation}
\label{Gselfb}
G^{self}_\textmd{b}(r,t)=\rho^1_\textmd{b}(r,t) +\rho^1_\textmd{s}(r,t),
\end{equation}
where both density profiles are obtained considering that there was a particle in the $\textmd{b}$-state at $r=0$ and $t=0$. $G^{self}_\textmd{b}(r,t)$ represents the probability density of finding the test particle  at distance $r$ from the origin at time $t$, given that at time $t=0$ it was located at $r=0$ in the $\textmd{b}$-state.

The corresponding MSD of the particles in the $\textmd{b}$-state is
\begin{equation}
\label{MSDb}
\langle \Delta r^2(t) \rangle_\textmd{b} = 4\pi \int_0^{\infty}r^4G^{self}_\textmd{b}(r,t)dr.
\end{equation}

The same procedure can be applied starting with a test particle in the $\textmd{s}$-state located at the origin to calculate $G^{self}_\textmd{s}(r,t)$ (defined as the probability density of finding the test particle at distance $r$ at time $t$, given that at time $t=0$ it was located at $r=0$ in the $\textmd{s}$-state) and $\langle \Delta r^2(t) \rangle_\textmd{s}$.

The average MSD of the system is 
\begin{equation}
\label{MSDave}
\langle \Delta r^2(t) \rangle_{ave}=x_\textmd{b}\langle \Delta r^2(t) \rangle_\textmd{b} + x_\textmd{s}\langle \Delta r^2(t) \rangle_\textmd{s} 
\end{equation}

Finally, the self-intermediate scattering function, routinely measured in light scattering experiments to characterize dynamics of colloidal systems, is 
\begin{equation}
F^{self}(\qq,t) = x_\textmd{s} F_\textmd{s}^{self}(\qq,t) + x_\textmd{b} F_\textmd{b}^{self}(\qq,t) \, .
\label{Fscat}
\end{equation}
where $F_i^{self}(\qq,t)$ are the partial self-intermediate scattering functions, obtained as the Fourier transform of the corresponding self-part of the van Hove distribution of displacements
\begin{equation}
F^{self}_i(q,t)=\frac{4\pi}{q}\int_0^{\infty}G_{i}^{self}(r,t)\sin(qr)rdr \ \ \ \ \ i=\textmd{b},\textmd{s}
\label{Fscat0}
\end{equation}

\section{An exactly solvable Continuous Time Random Walk model for active switching colloids}
\label{sec:CTRW}

In this section we describe a theoretical approach based on a Continuous Time Random Walk model (CTRW) to gain new insights on the complex diffusion of active switching colloids. In this approach the system is viewed as a heterogeneous assembly of big ($\textmd{b}$) and small ($\textmd{s}$) colloids which can stochastically switch from one state to another at constant rates. We are interested in particular in the distribution of particle displacements, i.e. the self-part of the van Hove distribution function $G^{self}(\rr,t)$ defined in Eq.~\ref{Gselfdef}, which measures the probability that a given colloid has traveled a distance $\rr$ in a time $t$. To account for the interplay between the colloids bare diffusion and their switching activity, we first define $g_{i}({\bf r},t)$ as the probability density function for a colloid to make a displacement $\rr$ in a time $t$, provided it is in state $i=\textmd{s,b}$ during the whole time interval $[0,t]$. In addition, let $p_{i}(t)$ be the probability that a colloid in state $i$ switches \emph{for the first time} to the complementary state $j$ at time $t$, and define the cumulative distribution
\be
P_{i}(t) \equiv \int_t^{\infty} p_{i}(t') dt' \, ,
\label{pcumul}
\ee
as the probability that the time to switch state ($i\to j$) is larger than $t$. In this way, a colloid that is initially (at time $t=0$) in state $i$ may persist in this state during a time $t$ with probability $P_{i}(t)$, or it may switch state $i\to j$ at some intermediate time $t'\in[0,t]$ with probability $p_{i}(t') dt'$. The stochastic state-switching activity $i\to j$ is assumed here to be homogeneous in time, with constant transition rates $k_{ij}$, which implies an exponential form for both $p_i(t)=k_{ij} \exp(-k_{ij}t)$ and $P_{i}(t)=\exp(-k_{ij}t)$. Moreover, as explained above, stationarity implies a detailed balance condition relating the transition rates between both states, namely $x_\textmd{s} k_{\textmd{sb}} = x_\textmd{b}k_{\textmd{bs}}$, with $x_i$ the fraction of colloids in state $i$ in the steady state.
 
Now, if $G^{self}_i(\rr,t)$ is the probability that a colloid \emph{initially in state $i$} travels a distance $\rr$ in a time $t$, we can write the following recurrence
\be
G_{i}^{self}({\bf r},t) = P_{i}(t) g_{i}({\bf r},t) + \int_0^t dt^{\prime} \int d{\bf r}^{\prime} p_{i}(t^{\prime}) g_{i}({\bf r}^{\prime},t^{\prime}) G_{j}^{self}({\bf r}-{\bf r}^{\prime},t-t^{\prime})
\label{propgel0}
\ee
or equivalently, in a more compact form,
\be
G_{i}^{self}({\bf r},t) =  P_{i}(t) g_{i}({\bf r},t) + \left[\Delta_{i} \circ G_{j}^{self}\right] (\rr,t) \, ,
\label{propgel}
\ee
where we defined $\Delta_i(\rr',t') \equiv p_i(t') g_i(\rr',t')$, and $\circ$ stands for spatio-temporal convolution. The first term describes the propagation of colloids which persist in the same state $i=\textmd{s,b}$ for the whole time interval, while the second term captures propagation in the presence of state changes at intermediate times. Recurrences like this one can be solved in general in the Laplace-Fourier domain~\cite{Hurtado2007}, where the integral equations boil down to simple algebraic relations for the transformed functions. In particular, defining the Laplace-Fourier transform of a generic function $h(\rr,t)$ as
\begin{equation}
h(\qq,s) \equiv \int_0^{\infty} dt \, \text{e}^{-s t} \int d\rr \,  \text{e}^{i \qq\cdot\rr} h(\rr,t) \, ,
\end{equation}
and applying this transformation to Eqs. \ref{propgel0}-\ref{propgel}, we obtain the following solution after some algebra,
\begin{equation}
G_{i}^{self}(\qq,s)=  \frac{\Delta_{i}(\qq,s)\left[k_{ij}^{-1}  + k_{ji}^{-1} \Delta_{j}(\qq,s)\right]}{1-\Delta_{i}(\qq,s) \Delta_{j}(\qq,s)} \, , \quad i=\textmd{s},\textmd{b}\, .
\label{FL}
\end{equation}
The colloids Brownian dynamics suggests to use diffusive expressions for the bare propagators,
\be
g_i(\rr,t)=\frac{\text{e}^{- \rr^2/(4D_i t)}}{(4\pi D_i t)^{3/2}} \, ,
\label{galpha}
\ee
with $D_i$ the colloid diffusion constant in state $i$. The Fourier-Laplace transform of $\Delta_i(\rr,t)$ is hence
\be
\Delta_i(\qq,s) = \frac{1}{1+k_{ij}^{-1} (s+ D_i \qq^2)} \, , 
\label{deltaA}
\ee
which in turn leads to
\be
G_{i}^{self}(\qq,s)=  \frac{s + k_{ij} + k_{ji} + D_j \qq^2}{(s+k_{ij} + D_i \qq^2)(s+k_{ji} + D_j \qq^2) - k_{ij} k_{ji}} \, , 
\label{GGalpha}
\ee
with $i=\textmd{s},\textmd{b}$ and $j=\textmd{b},\textmd{s}$ complementary to $i$. Note that the Fourier-Laplace transform of the self-part of the van Hove distribution function can be written in general as $G^{self}(\qq,t) = x_\textmd{s} G_\textmd{s}^{self}(\qq,t) + x_\textmd{b} G_\textmd{b}^{self}(\qq,t)$. Before proceeding, let us analyze the long-time, large-lengthscale asymptotic behavior predicted by Eq. \ref{GGalpha}. In particular, taking the limit $s\to 0$, $\qq\to 0$ in this expression, with $s/\qq^2\to \text{constant}$, we obtain a standard diffusive propagator at leading order, i.e. 
\be
G_{i}^{self}(\qq,s) \xrightarrow[\qq^2\to 0]{s\to 0} \frac{1}{s + D_{ave} \qq^2} \, ,
\label{Gave}
\ee
with an effective, average diffusion constant
\be
D_{ave} = \frac{k_{\textmd{sb}} D_\textmd{b} + k_{\textmd{bs}} D_\textmd{s}}{k_{\textmd{sb}}+k_{\textmd{bs}}} = x_\textmd{s} D_\textmd{s} + x_\textmd{b} D_\textmd{b} \, ,
\label{Dave}
\ee
where we have used the detailed balance condition $x_\textmd{s} k_{\textmd{sb}} = x_\textmd{b} k_{\textmd{bs}}$ in the last equality. Eq.~\ref{Gave} is nothing but the Fourier-Laplace transform of a Gaussian (diffusive) displacement distribution (akin to Eq.~\ref{galpha}) with effective diffusion constant $D_{ave}$, and is independent of the initial state of the colloid, as expected.

Inverting now the Laplace transform in Eq.~\ref{GGalpha}, we obtain the partial self-intermediate scattering function of particles initially in the $i$-state
\ba
\label{Fs_CTRW}
F_i^{self}(\qq,t) &=& \text{e}^{-t\Theta(\qq)/2} \bigg\{ 
\cosh \left[\frac{t}{2} \sqrt{\Lambda(\qq)} \right]  \label{vanH}
\\
&+& \frac{2\Gamma_i(\qq) - \Theta(\qq)}
{\sqrt{\Lambda(\qq)}} \sinh \left[\frac{t}{2} \sqrt{\Lambda(\qq)}\right] \bigg\} \, , \nonumber
\ea
where 
\ba
\Gamma_i(\qq) & = & k_{ij} + k_{ji}+ D_j\qq^2 \, . \label{Gamma}\nonumber\\[-2pt]
\Theta(\qq) & = & k_{ij} + k_{ji}+ (D_i + D_j)\qq^2 \, , \label{Theta} \nonumber\\[-2pt] 
\Lambda(\qq) & = & (k_{ij} + k_{ji})^2 + 2 (k_{ij} - k_{ji})(D_i - D_j) \qq^2 + (D_i - D_j)^2 \qq^4 .\label{Lambda}\nonumber 
\ea 
The full scattering function is obtained from Eq.~\ref{Fscat}. From this expression we can further compute analytically the average MSD, $\la \Delta \rr^2(t)\ra_{ave}$, as
\be
\hspace{-0.2cm}\la \Delta \rr^2(t)\ra_{ave} = -{\frac{\partial^3 \left[q F^{self}(\qq,t) \right]}{\partial q^3}}\bigg |_{\qq=0} = x_\textmd{s} \la \Delta \rr^2(t)\ra_\textmd{s} + x_\textmd{b} \la \Delta \rr^2(t)\ra_\textmd{b}
\label{MSD1}
\ee
where $\la \Delta \rr^2(t)\ra_i$ is the MSD for a colloid in state $i=\textmd{s,b}$ at time $t=0$. Differentiating now the exact expression Eq. \ref{vanH}, we obtain
\ba
\label{MSD_CTRW}
\la \Delta \rr^2(t)\ra_i &=& \frac{6}{(k_{ij} + k_{ji})^2} \left[ k_{ij}(D_i-D_j)(1-\text{e}^{-(k_{ij} + k_{ji})t}) \right. \nonumber \\
&+& \left. (k_{ij} + k_{ji})(D_i k_{ji} + D_j k_{ij}) t\right] \, .
\ea

It is important to emphasize here that the CTRW approach neglects particle-particle interactions, so it is expected to provide good predictions for the single particle dynamics only for diluted or weakly interacting colloidal systems. For strongly interacting systems, the CTRW theory could still be applied assuming that the values of $D_\textmd{b}$ and $D_\textmd{s}$ are given by the effective long-time concentration-dependent diffusion coefficients.


\section{Reactive Brownian dynamics (R-BD) computer simulations}

In addition to the R-DDFT calculation and the CTRW theory described above, we performed reactive Brownian Dynamics (R-BD) for all systems as in our previous work~\cite{Bley2021}. R-BD allows a direct access to the dynamics and diffusion coefficients for validating the results of our R-DDFT while bypassing the extensive calculations of the self and distinct part of the van Hove function. The equation of motion for a particle $i$ in the particle-resolved simulations using the overdamped Langevin equation is written as
\begin{equation}
\xi_{i} \mathbf{\dot{r}}_{i} = - \nabla U(\mathbf{r}_{i}) + \mathbf{R}(t) \; ,
\end{equation} 
where $\mathbf{\dot{r}}_{i}$ and $\mathbf{r}_{i}$ denote the velocity and the position vector of the $i$-th particle, $\mathbf{R}(t)$ is a random force vector, and the drag coefficient $\xi_{i}$ is related with the diffusion coefficient $D_{i}$ through $D_{i} = k_{\mathrm{B}}T/\xi_{i}$. Following the fluctuation-dissipation theorem, the components of the random force vector fulfill the properties $\langle R_{\alpha}(t)\rangle = 0$ and $\langle R_{\alpha}(t) R_{\beta}(t') \rangle = 2 \xi_{i}^{2} D_{i} \delta_{\alpha \beta} \delta (t - t^{\prime})$ with $\alpha$ and $\beta$ representing the spatial dimensions, $\delta_{\alpha \beta}$ the Kronecker delta, and $\delta(t-t^{\prime})$ denoting the Dirac delta function. In the absence of an external field, the force vector acting on each of the $N$ particles $\mathbf{F}_{i}$ arises only from the pairwise, distance-dependent interactions $u_{ij}(r_{ij})$ following Eq.~\ref{ugaussian}, and thus the force writes $\mathbf{F}_{i} = -\nabla U(\mathbf{r}_{i}) = - \sum_{i \neq j}^{N} \nabla u_{ij}(r_{ij})$. All particle positions are updated after each time interval $\Delta t$ using the Euler-Maruyama propagation scheme \cite{Ermak1978}, that is
\begin{equation}
\mathbf{r}_{i}(t + \Delta t) = \mathbf{r}_{i}(t) + \frac{\Delta t}{\xi_{i}} \mathbf{F}_{i} + \sqrt{2D_{i} \Delta t} \pmb{\zeta}_{i} \ , 
\end{equation}
where $\Delta t = 10^{-4} \tau_{\mathrm{B}}$ as the integration time-step and $\pmb{\zeta}_{i}$ is a vector consisting of random values following a standard normal distribution. The two probabilities of switching $p_{\mathrm{bs}}$ and $p_{\mathrm{sb}}$, which check for switching events every integration step $\Delta t$, are
\begin{equation}
p_{\mathrm{bs}} = 1 - e^{-k_{\mathrm{bs}}\Delta t}, \ \ \ p_{\mathrm{sb}} = 1 - e^{-k_{\mathrm{sb}}\Delta t} \; . 
\end{equation}
The properties of the $i$-th particle are switched if a random variate following a uniform distribution between zero and one is below $p_{\mathrm{sb}}$ if the particle is at state $\textmd{s}$, or is below $p_{\mathrm{bs}}$ if the the particle is at state $\textmd{b}$. Our R-BD simulations for up to 500 switching particles have been conducting using an own code for production time up to $50 \tau_{\mathrm{B}}$ and system S2 presented in Table \ref{tbl:systems}. For each activity, we collected five different trajectories from simulations of cubic cells of edge length $12.8 \sigma_{\mathrm{s}}$ with periodic boundary conditions. For accessing the diffusion coefficients $D_{\mathrm{b}}$ and $D_{\mathrm{s}}$, the MSD for the corresponding species is calculated with respect to the initial positions $r_{\mathrm{i}}(0)$ and states, and is written as
\begin{equation}
\langle \Delta r^{2}(t) \rangle_{i} = \langle (\mathbf{r}(t) - \mathbf{r}(0))^{2} \rangle_{i} = \frac{1}{N_{i}} \sum_{\mu = 0}^{N_{i}} (\mathbf{r}_{\mu}(t) - \mathbf{r}_{\mu}(0))^{2} \ \ \ \ i=\textmd{b},\textmd{s} 
\end{equation}
where $N_{i}$ is the corresponding amount of particles in state $i$, and $\mathbf{r}_{\mu}(t)$ and $\mathbf{r}_{\mu}(0)$ denoting the position vectors of each particle in state $i$ at time $t$ and at the initial state.

\section{Results and discussion}

We apply our generalized R-DDFT method to deduce the particle dynamics of two representative systems, S1 and S2, specified in Table~\ref{tbl:systems}. The interaction parameters and total particle bulk concentration for both systems are exactly the same, but the composition ($x_\textmd{s}$) and the diffusion coefficient of the particles in the $\textmd{b}$-state ($D_\textmd{b}$) are different. In particular, for system S1 the particle diffusivities follow the Stokes-Einstein relation, $D_\textmd{b}=(\sigma_\textmd{s}/\sigma_\textmd{b})D_\textmd{s}=0.5D_\textmd{s}$, whereas for system S2 both particles have very dissimilar diffusivities, namely $D_\textmd{b}=0.01D_\textmd{s}$. In both systems, the kinetic rate constants fulfill the condition given by Eq.~\ref{condition}, i.e. $k_\textmd{bs}/k_\textmd{sb}=x_\textmd{s}/x_\textmd{b}$, in order to preserve the relative bulk composition of colloids in $\textmd{b}$ and $\textmd{s}$ states.
\begin{table}[h]
	\small
	\caption{\ Main parameters describing the particle interactions and concentrations for two different active switching Gaussian colloidal systems.}
	\label{tbl:systems}
	\begin{tabular*}{0.48\textwidth}{@{\extracolsep{\fill}}lllllllll}
		\hline
		System & $\epsilon_\textmd{bb}$ & $\epsilon_\textmd{ss}$ & $\epsilon_\textmd{bs}$ & $\sigma_\textmd{b}/\sigma_\textmd{s}$ &  $\sigma_\textmd{bs}/\sigma_\textmd{s}$ & $\rho_\textmd{T}\sigma_\textmd{s}^3$ & $x_\textmd{s}$ & $D_\textmd{b}/D_\textmd{s}$ \\
		\hline
		S1 & 2.0 & 2.0 & 2.0 & 2.0 & 1.5 & 0.239 & 0.8 & 0.5 \\
		S2 & 2.0 & 2.0 & 2.0 & 2.0 & 1.5 & 0.239 & 0.2 & 0.01 \\
		\hline
	\end{tabular*}
\end{table}

\subsection{Results for system S1}

The calculation are initiated assuming that the distinct parts of the van Hove function at time $t=0$, $\rho^2_{i}(t,t=0)$, are given by the corresponding steady-states radial distribution functions of the active colloidal system (Eqs.~\ref{initial1} if the test central particle is in the $\textmd{b}$-state, or \ref{initial2} if it is in the $\textmd{s}$-state). For this purpose, we first need to solve the 4-state R-DDFT equations to calculate the steady-state radial distribution functions of the active system, $g_{ij}(r)$, obtained in the limit $t \rightarrow \infty$~\cite{MonchoJorda2020,Bley2021}. We emphasize again that the steady-state distributions only agree with the equilibrium distributions $g^{eq}_{ij}(r)$ for $a=0$. Otherwise (for $a>0$) the steady state corresponds to a time-independent non-equilibrium configuration.
\begin{figure*}[ht!]
	\centering
	\includegraphics[width=1.0\linewidth]{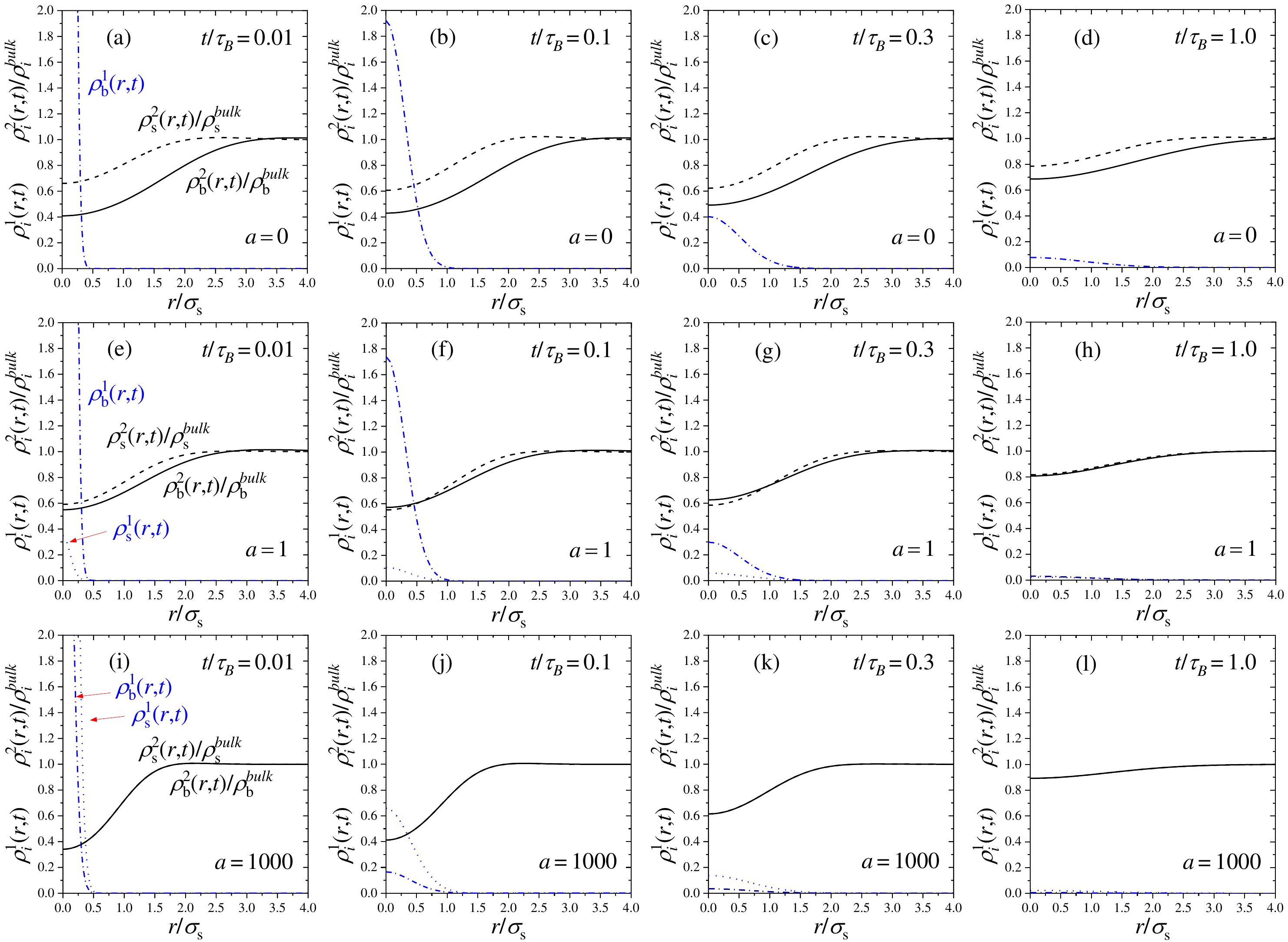}
	\caption{Particle densities $\rho^1_\textmd{b}(r,t)$ and $\rho^1_\textmd{s}(r,t)$ (blue lines) and $\rho^2_\textmd{b}(r,t)$ and $\rho^2_\textmd{s}(r,t)$ (black lines) for an active switching fluid of Gaussian colloids as a function of the scaled distance $r/\sigma_\textmd{s}$ at different times, obtained for system S1. Plots depict the results for three activity rates ($a=0$, $1$ and $1000$) and four times, $t/\tau_B=0.01$, $0.1$, $0.3$ and $1$, respectively.}
	\label{fig:VanHove2_rhos}
\end{figure*}

For the self-part, a very short initial time $t=t_0=5\times 10^{-5}\tau_B$ is used to approximate the singular $\delta$-function of $\rho^1_\textmd{i}(r,t=0)$ by a very sharp Gaussian distribution
\begin{equation}
\rho^1_i(r,t_0)=\frac{1}{(4\pi D_\textmd{s} t_0)^{3/2}}e^{-r^2/(4D_\textmd{s}t_0)} \ \ \ i=\textmd{b},\textmd{s}
\end{equation}
where $t_0$ is small enough to assure that the distinct parts have barely changed from the initial distribution.

Fig.~\ref{fig:VanHove2_rhos} shows the time dependent density profiles $\rho^{\alpha}_i(r,t)$ for system S1 of Table~\ref{tbl:systems} obtained assuming that at time $t=0$ there was a particle in the $\textmd{b}$-state located in the origin. The twelve plots represent the results for three different activity rates ($a=0$, $1$ and $1000$) and four times ($t/\tau_B=0.01$, $0.1$, $0.3$ and $1$). All times are normalized by the Brownian time, $\tau_B=\sigma_\textmd{s}^2/D_\textmd{s}$.

We first examine the plots for $a=0$, which correspond to a non-active equilibrium mixture of non-switching big and small Gaussian colloids (see Figs.~\ref{fig:VanHove2_rhos}(a)-(d)). Since switching is forbidden, the central big test particle remains in the $\textmd{b}$-state all the time during the diffusion process. The time evolution of the self and distinct part are very similar to the ones observed in one-component systems~\cite{Archer2007}. The initial delta peak of the self part, $\rho^1_b(r,t)$, broadens as time increases. Conversely, the distinct parts providing the density profiles of surrounding particles in the $\textmd{b}$ and $\textmd{s}$ states tend to be more uniform. Indeed, the initial depletion region of $\rho^{2}_i(r,t)$ close to the origin is progressively reduced, and becomes flatter as time evolve. In the limit $t\rightarrow \infty$ or $r\rightarrow \infty$, the self part tends to zero whereas the distinct part converges to the uniform distribution.
\begin{figure*}[ht!]
	\centering
	\includegraphics[width=1.0\linewidth]{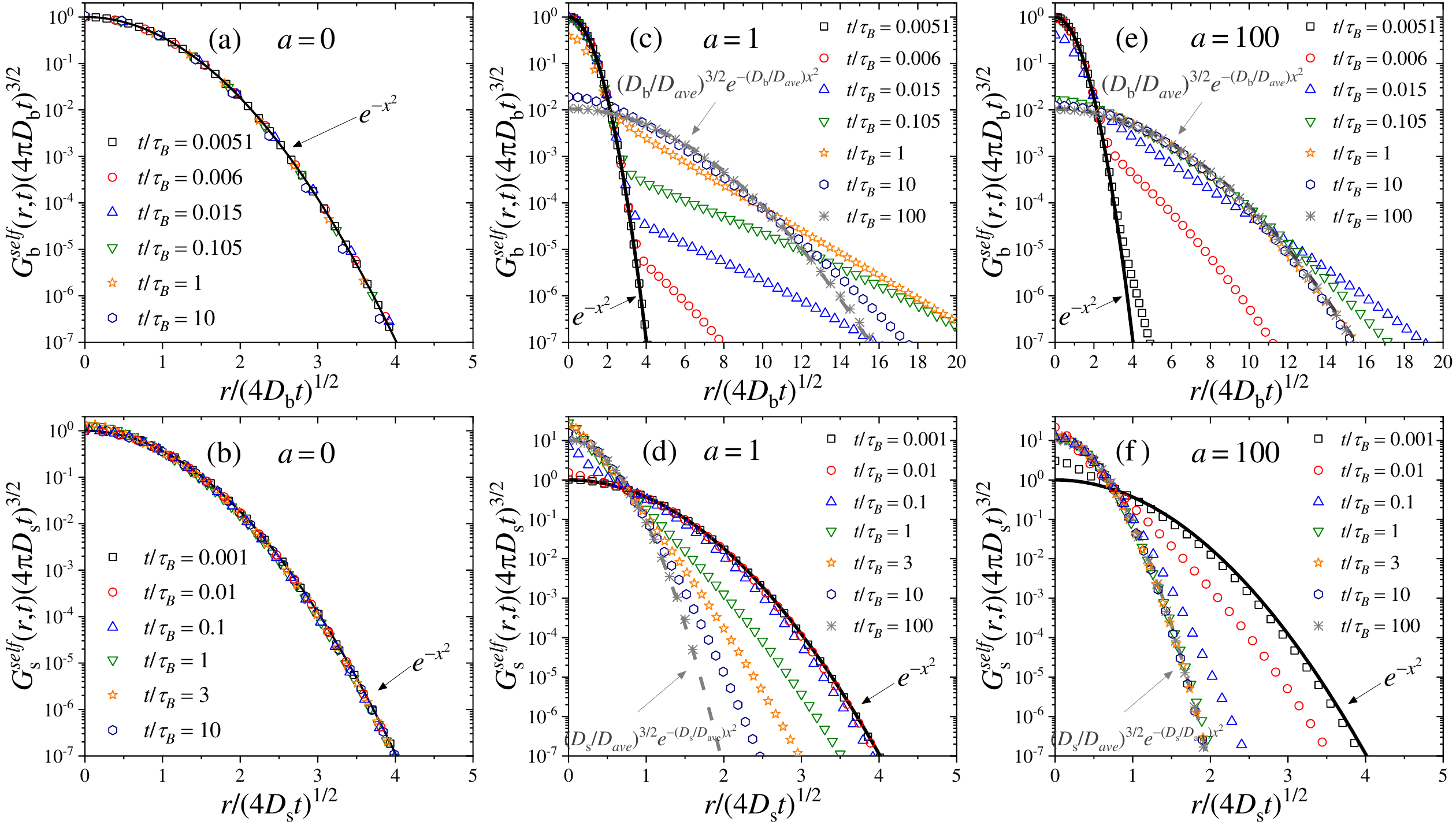}
	\caption{Plots (a), (c) and (e): Scaled self-part of the van Hove distribution of displacements for a particle initially in the $\textmd{b}$-state, $G^{self}_\textmd{b}(r,t)(4\pi D_\textmd{b} t)^{3/2}$, as a function of the scaled distance $r/(4D_\textmd{b}t)^{1/2}$, for $a=0$, $1$ and $100$. Plots (b), (d) and (f): Scaled self-part of the van Hove function for a particle initially in the $\textmd{s}$-state , $G^{self}_\textmd{s}(r,t)(4\pi D_\textmd{s} t)^{3/2}$ as a function of the scaled distance $r/(4D_\textmd{s}t)^{1/2}$, for $a=0$, $1$ and $100$. Calculations performed for system S2.}
	\label{fig:VanHove2_Gself_bs_full}
\end{figure*}

The situation changes if we consider active switching systems ($a>0$). In this case, particles not only diffuse, but they are also able to switch between states $\textmd{b}$ and $\textmd{s}$. This means that, if for instance the system starts with a test big particle at $t=0$ (so $\rho^1_\textmd{b}(r,0)=\delta(\mathbf{r})$ and $\rho^1_\textmd{s}(r,0)=0$), for larger times some fraction of this initial distribution broadens by diffusion, but other part switches to create particles in the $\textmd{s}$-state, leading to time-dependent density distributions for particles in both states, $\rho^1_\textmd{b}(r,t)$ and $\rho^1_\textmd{s}(r,t)$. This effect, which constitutes a new exclusive feature of active switching colloids, can be clearly observed in Fig.~\ref{fig:VanHove2_rhos}. Indeed, for $a>0$, the initial $\delta$-distribution of $\rho^1_\textmd{b}(r,t)$ not only spreads, but also leads to the formation of a new distribution of small colloids, $\rho^1_\textmd{s}(r,t)$, indicating that the central particle has switched from the $\textmd{b}$-state to the $\textmd{s}$-state. As shown by Figs.~\ref{fig:VanHove2_rhos}(e)-(h), this effect is small for $a=1$ at times below $t/\tau_B = 1$, but it becomes more important for longer times. Conversely, For $a=1000$ (Figs.~\ref{fig:VanHove2_rhos}(i)-(l)), the switching rate is so fast that the time evolution generates a large peak of particles in the $\textmd{s}$-state at very short times. We have shown in our earlier works that, in the limit of a fast switching rate, the switching events are much faster than the diffusion, so the system behaves as an effective one-component system in which all particles interact with the same effective pair potential~\cite{MonchoJorda2020,Bley2021}. As a consequence, both density profiles rapidly converge to a common shape, with $\rho^1_\textmd{s}(r,t)=(x_\textmd{s}/x_\textmd{b})\rho^1_\textmd{b}(r,t)$. The same happens for the distinct part of the van Hove function, $\rho^2_\textmd{s}(r,t)=(x_\textmd{s}/x_\textmd{b})\rho^2_\textmd{b}(r,t)$.

\subsection{Results for the dynamically more asymmetric system S2}

In order to really understand the role of the diffusion coefficients on the dynamic behavior of the active system, we have performed the same calculations, but using system S2 shown in Table~\ref{tbl:systems}. System S2 has similar features than system S1, but with a very important difference: the diffusion coefficient of colloids in the $\textmd{b}$-state is chosen to be very small compared to the one for particles in the $\textmd{s}$-state, $D_\textmd{b}=0.01D_\textmd{s}$, resembling the features of systems with highly heterogeneous dynamics~\cite{Hurtado2007}. We calculate the self-part of the van Hove function $G^{self}_\textmd{b}(r,t)$ (see Eq.~\ref{Gselfb}), starting with a the test particle in the $\textmd{b}$-state located at $\mathbf{r}=0$ at time $t=0$. The same calculation is repeated choosing a test particle initially in the $\textmd{s}$-state to determine $G^{self}_\textmd{s}(r,t)$.

Fig.~\ref{fig:VanHove2_Gself_bs_full}(a) shows $G^{self}_\textmd{b}(r,t)$ obtained for a binary colloidal system S2 in equilibrium ($a=0$). In particular, we plot the scaled functions $(4\pi D_\textmd{b}t)^{3/2}G^{self}_\textmd{b}(r,t)$ against the scaled distance $x=r/(4D_\textmd{b}t)$. As observed, all the curves for different times collapse in a common form given by $e^{-x^2}$, which indicates that the diffusion process of the test big particle follows a Gaussian distribution with a diffusion coefficient given by $D_\textmd{b}$. The same conclusion is found if we consider a test particle in the $\textmd{s}$-state and plot the scaled $(4\pi D_\textmd{s}t)^{3/2}G^{self}_\textmd{s}(r,t)$ against $r/(4D_\textmd{s}t)$, as shown in Fig.~\ref{fig:VanHove2_Gself_bs_full}(b). Therefore, we conclude that the time-dependent distribution of displacement of the particles in a non-active ($a=0$) equilibrium fluid mixture of interacting Gaussian colloids obeys the well-known Gaussian distribution
\begin{equation}
\label{GGaussian}
G^{self}_\textmd{i}(r,t)=\frac{1}{(4\pi D t)^{3/2}}e^{-r^2/(4Dt)} 
\end{equation}
where $D$ is the corresponding diffusion coefficient of specie $i$, namely $D=D_i$ (with $i=\textmd{b},\textmd{s}$).

\begin{figure*}[ht!]
	\centering
	\includegraphics[width=1.0\linewidth]{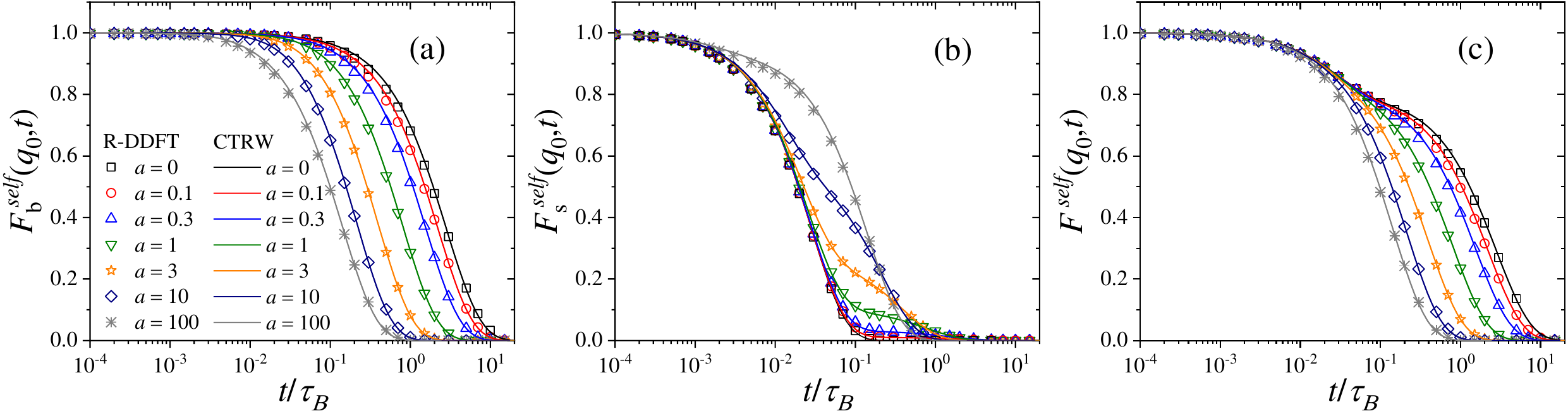}
	\caption{Partial and full self-intermediate scattering functions as a function of time for different activity rates, from $a=0$ to $a=100$, calculated for $q_0\sigma_s=6$: (a) $F^{self}_\textmd{b}(q_0,t)$, (b) $F^{self}_\textmd{s}(q_0,t)$ and (c) $F^{self}(q_0,t)=x_\textmd{b}F^{self}_\textmd{b}(q_0,t)+x_\textmd{s}F^{self}_\textmd{s}(q_0,t)$. Calculations performed for system S2.}
	\label{fig:Fs_q6}
\end{figure*}

This scaling behavior breaks down for active switching system. Figs.~\ref{fig:VanHove2_Gself_bs_full}(c) and (d) illustrate exactly the same scaled functions for system S2 at several times, but turning on the non-equilibrium switching activity rate at $a=1$. Fig.~\ref{fig:VanHove2_Gself_bs_full}(c) corresponds to the self-part of the van Hove function for a test particle in the $\textmd{b}$-state at $t=0$. For very short times, $t\ll \tau_B/a$, the big test particle still preserves its identity, so it follows the Gaussian distribution of displacements given by Eq.~\ref{GGaussian}, with a diffusion coefficient $D=D_\textmd{b}$. However, for intermediate times, $t \sim \tau_B/a$, $G^{self}_\textmd{b}(r,t)$ is not Gaussian any more. In fact, it follows different distributions for short and large displacements, showing a non-Gaussian tail with a well-defined shoulder that represents the crossover from one behavior to the other. In particular, large displacements have a much larger probability to occur compared to the Gaussian prediction. This is due to the fact that the original test particle in the $\textmd{b}$ has already switched to a faster $\textmd{s}$-state. Clearly, $G^{self}_\textmd{b}(r,t)$ exhibits a bimodal character suggesting the existence of slow particles and faster diffusing particles. This dynamic behavior resembles the one observed in heterogeneous diffusion of reversible attractive colloidal gels, in which the system is formed by a coexistence of slow percolating cluster of connected droplets and fast, more freely diffusing droplets, with a dynamic exchange between the two families set by polymer moves~\cite{Hurtado2007}. In our case, this lack of Gaussianity is a clear signature of the non-equilibrium activity introduced by the particle switching. Note that similar heterogeneous dynamics has been also observed in a wide variety of systems with Brownian yet non-Gaussian diffusion \cite{Chechkin2017,Pastore2021,Miotto2021}.

For very long times, ($ t\gg \tau_B/a$), the initial big test particle has experienced many switching events between $\textmd{b}$ and $\textmd{s}$, and vice versa. In this limit, $G^{self}_\textmd{b}(r,t)$ recovers again the Gaussian behavior (cf. Eq.~\ref{GGaussian}), but with an average diffusion coefficient given by the mean of the individual diffusion coefficients weighted by the kinetic rate constants, given by Eq.~\ref{Dave}. This is exactly the prediction derived within the CTRW theory described in Section~\ref{sec:CTRW} for the self-part of the van Hove distribution in the long-time large-lengthscale limit.

A similar behavior is found if we analyze the self-part of the distribution of displacements for a test particle in the $\textmd{s}$-state at $t=0$ (see Fig.~\ref{fig:VanHove2_Gself_bs_full}(d)). For short times ($t\ll \tau_B/a$), the small test particle has not yet experienced any switching event, so $G^{self}_\textmd{s}(r,t)$ follows a Gaussian distribution with $D=D_\textmd{s}$. At intermediate times, the Gaussian behavior is lost. In this case, large displacements have smaller probability to occur compared to the Gaussian dependence, indicating that the test particle is now diffusing slower due to the switching from $\textmd{s}$ to $\textmd{b}$-state. For large enough elapsed times, the occurrence of multiple switching events between both particle states finally leads to a new Gaussian distribution with $D=D_{ave}$.

The same transient behavior is found increasing the activity rate to $a=100$. The only difference is that, in this case, switching events befall $100$ times faster, so the transition from the short-time Gaussian regime to the long-time one arises at much shorter times. This effect can be clearly observed in Figs.~\ref{fig:VanHove2_Gself_bs_full}(e) and (f), where $G^{self}_i(r,t)$ becomes non-Gaussian for $t \ll \tau_B$. 

\subsection{Intermediate scattering function}

An observable of direct experimental relevance is the self-intermediate scattering function $F^{self}(q,t)$, as well as its partial components $F^{self}_i(q,t)$ ($i=$s,b) defined in Eq.~\ref{Fscat0}. These functions characterize the system dynamical relaxation at the single-particle level for different wave vectors $\qq$, or equivalently at lengthscales inversely proportional to $|\qq|$. Fig.~\ref{fig:Fs_q6} shows $F^{self}(q,t)$ as well as $F^{self}_\textmd{b}(q,t)$ and $F^{self}_\textmd{s}(q,t)$ for $q\sigma_\text{s}=6$ and system S2, as obtained within the R-DDFT approach together with the CTRW prediction (Eqs.~\ref{Fscat}, \ref{Fscat0} and \ref{Fs_CTRW}). Focusing first on the small (s) colloids, note that they exhibit a fast initial relaxation, as captured by the quick initial decay of $F^{self}_\textmd{s}(q,t)$, see Fig.~\ref{fig:Fs_q6}(b), on a time scale related to their relatively large diffusion constant $D_\text{s}$. However, the active state switching eventually kicks in, triggering a slower relaxation for initially small particles which have switched to the big (b) state, with much smaller diffusion constant $D_\textmd{b}=10^{-2} D_\textmd{s}$. This gives rise to a plateau in $F^{self}_\textmd{s}(q,t)$ at intermediate timescales. Interestingly, the height of the plateau decreases with decreasing activity $a$. Indeed, for small activity the time between switching events $\text{s}\to\text{b}$ is large, so most small colloids have time to diffusively relax in the medium before a switching event slows down their dynamics, leading to small plateau height. Conversely, for large activities $a\gg1$ very few small colloids have time to relax before switching to the big (slow) state, and hence the amplitude of the plateau is large in this case. On the other hand, for big (b) colloids $F^{self}_\textmd{b}(q,t)$ exhibits a simpler relaxation pattern, see Fig.~\ref{fig:Fs_q6}(a). In this case the initial relaxation of big colloids is already slow, as expected from their low bare diffusivity, and this relaxation can be only accelerated by switching events to the small (s) state. In this way $F^{self}_\textmd{b}(q,t)$ exhibits no secondary relaxation plateau, and the relaxation timescale for b-colloids decreases with increasing activity. The combined relaxation of small and big active colloids gives rise to a typical two-step relaxation pattern for the global incoherent scattering function $F^{self}(q,t)=x_\textmd{b}F^{self}_\textmd{b}(q,t)+x_\textmd{s}F^{self}_\textmd{s}(q,t)$, see Fig.~\ref{fig:Fs_q6}(c), with a plateau at intermediate timescales with a height proportional to $x_\textmd{b}=1-x_\text{s}$, the fraction of big (i.e. slow) colloids in the stationary suspension. The relaxation timescale to the plateau is controlled by the small colloids fast bare diffusivity, while the global relaxation timescale decreases with increasing activity, as expected.

The lines in Fig.~\ref{fig:Fs_q6} correspond to the CTRW prediction given by Eq.~\ref{Fs_CTRW}, and the agreement with R-DDFT results is excellent in all cases, capturing in full detail the complex relaxation dynamics of the active colloidal suspension. This striking agreement (also observed below for MSDs and other dynamical quantities) strongly suggests that the soft colloidal interactions seem to play no significant role in the diffusive and local relaxation properties of the active suspension, at least for the interaction parameters explored in this paper. We expect interactions to become more important for larger values of $\epsilon_{ij}$, as suggested in Refs.~\cite{Mausbach2006,Wensink2008}. These interactions can be however taken into account within the CTRW model here introduced via renormalized diffusive propagators.

\subsection{Mean squared displacements (MSDs) and R-BD trajectories}

\begin{figure}[ht!]
	\centering
	\includegraphics[width=0.9\linewidth]{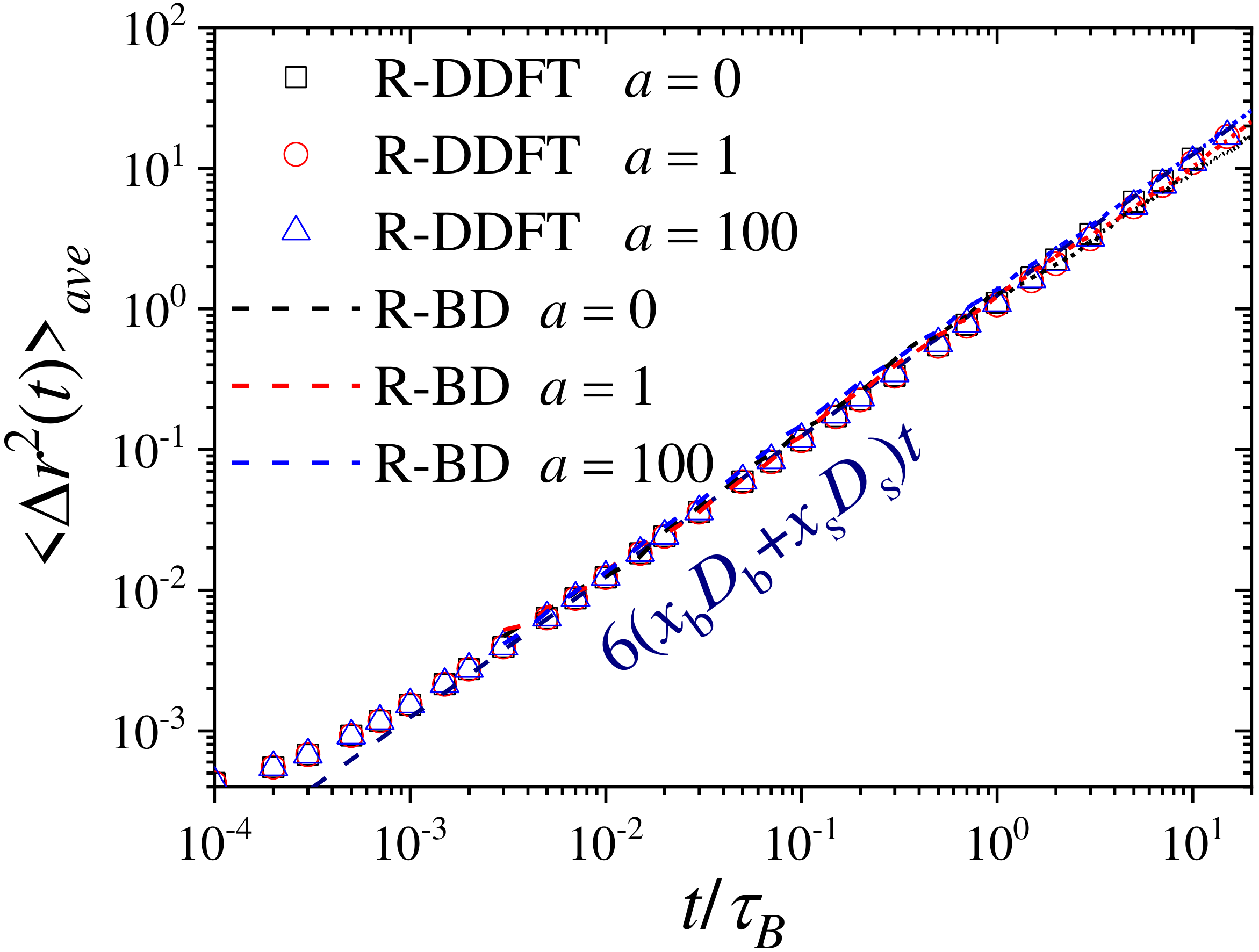}
	\caption{Average MSD of the active switching system as a function of time for system S2 for $a=0$, $1$ and $100$. Symbols are R-DDFT predictions, whereas dashed lines denote R-BD results.}
	\label{fig:VanHove2_MSDave}
\end{figure}

The relaxation dynamics and diffusive behavior at short and long timescales can be further understood by analyzing the MSD of the colloids. Fig.~\ref{fig:VanHove2_MSDave} shows the average MSD of system S2, $\langle \Delta r^2(t) \rangle_{ave}$ (calculated from Eq.~\ref{MSDave}) for $a=0$, $1$ and $1000$ obtained from the R-DDFT predictions and from R-BD simulations. Interestingly, theoretical and simulation data of the global MSD for these three activity rates collapse after a short transient in the same common linear dependence, given by $\langle \Delta r^2(t) \rangle_{ave}=6D_{ave}t$, with $D_{ave}=x_\textmd{b}D_\textmd{b}+x_\textmd{s}D_\textmd{s}$ being the average diffusion coefficient of the active system. In other words, from a global point of view, the system dynamics is Brownian and can be apparently modeled by a single average diffusion constant that is exactly the same regardless the activity rate, $a$. This is however in stark contrast with the non-Gaussian character of the diffusive dynamics captured by the self-part of the global and partial van Hove distributions of displacements (see Fig.~\ref{fig:VanHove2_Gself_bs_full} and the associated discussion). Therefore, the active colloidal mixture here studied exhibits the hallmarks of the \emph{Brownian yet non-Gaussian diffusion} phenomenon already described in other heterogeneous materials \cite{Chechkin2017,Pastore2021,Miotto2021}.

\begin{figure}[ht!]
	\centering
	\includegraphics[width=0.9\linewidth]{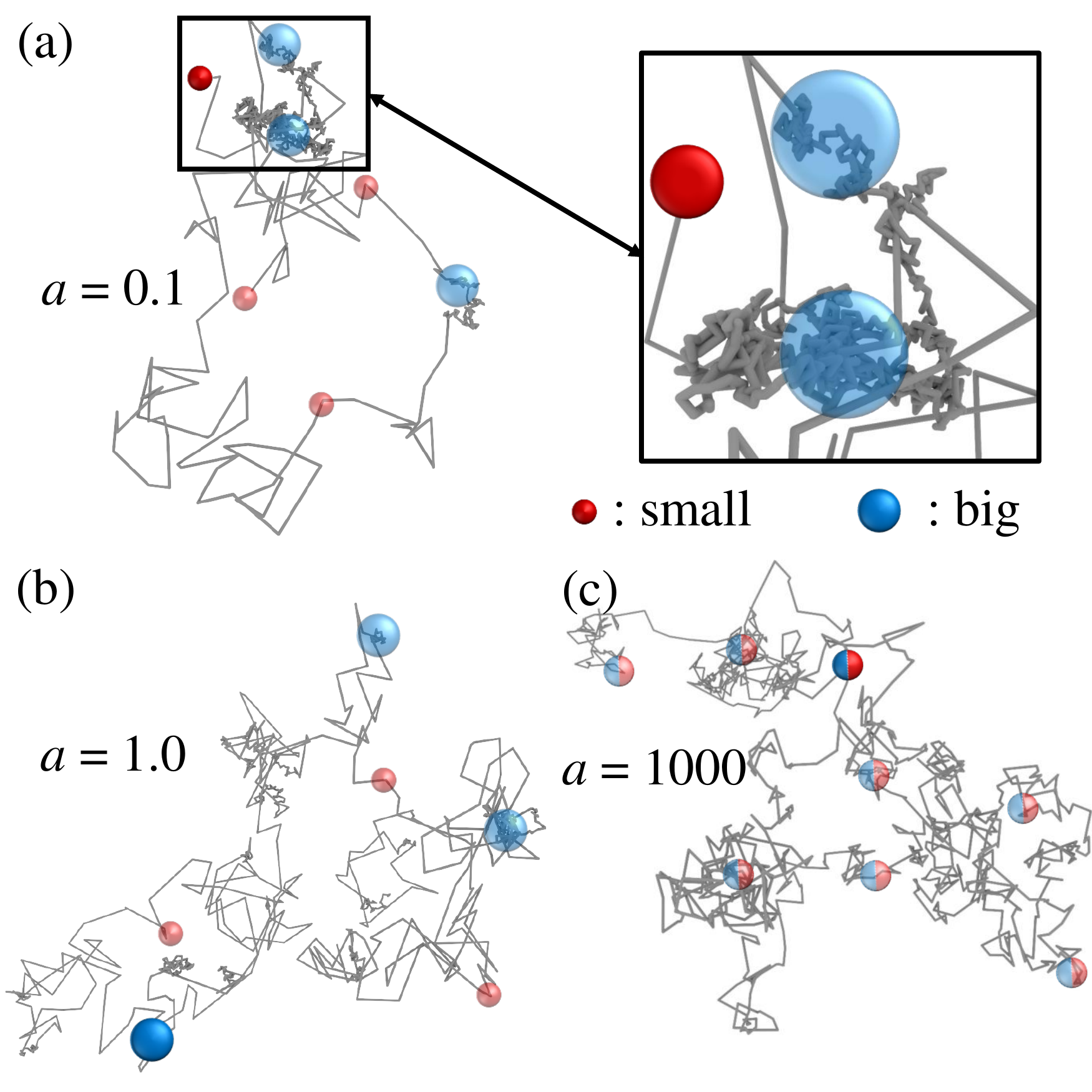}
	\caption{Trajectories obtained from R-BD simulations (gray lines) of single particles actively switching between their big (blue) and small (red) state at activities (a) $a = 0.1$ with a maximized region of slow diffusion being in the big state, (b) $a = 1.0$, and (c) $a = 1000$, where the split coloring indicates fast switching between the two states. Transparent colors represent past states and positions, non-transparent coloring indicates the final position and state. All trajectories cover a total time of $25~\tau_{\mathrm{B}}$ with a resolution of $0.025~\tau_{\mathrm{B}}$ for the displacement steps (system S2).}
	\label{fig:Particle_Trajectories}
\end{figure}

To better understand this phenomenon, we explore now some representative single-particle trajectories. In particular, Fig.~\ref{fig:Particle_Trajectories} contains a set of selected trajectories for a single, actively switching particle at different switching activities $a$. The lower $a$ is, the longer the particle stays in a given state $\textmd{b}$ or $\textmd{s}$, and thus the larger the explored regions with the corresponding diffusion coefficients $D_{\mathrm{s}}$ or $D_{\mathrm{b}}$. Since $D_{\mathrm{b}} \ll D_{\mathrm{s}}$, big particles diffusively explore relatively small volumes (see Fig.~\ref{fig:Particle_Trajectories}(a)), leading to local quasi-arrested dynamics, while small particles can travel further away exploring larger volumes for the same time interval.

For small activities, or equivalently large switching time intervals, the dynamics is clearly intermittent and heterogeneous, characterized by strong diffusion intervals punctuated with quasi-arrested periods when colloids switch to the big state, resulting in general in a highly heterogeneous distribution of particle displacements. When the activity increases, see Figure~\ref{fig:Particle_Trajectories}(b) for $a = 1.0$, the dynamical heterogeneity of the trajectory is less apparent as state switching events are more frequent and regions of fast and slow diffusion cannot be clearly delimited, leading to a more homogeneous distribution of displacement vectors. At much larger switching activities ($a =1000$, Fig.~\ref{fig:Particle_Trajectories}(c)), the switching intervals become sufficiently small to show a behaviour close to an effective one component (EOC) system~\cite{Bley2021}, where the macroscopic characterization of our systems revealed almost indistinguishable properties for big and small particles. However, here we still see a non-uniform distribution of displacements showing that the states remain distinguishable on a microscopic level for a single particle.

\begin{figure}[ht!]
	\centering
	\includegraphics[width=0.9\linewidth]{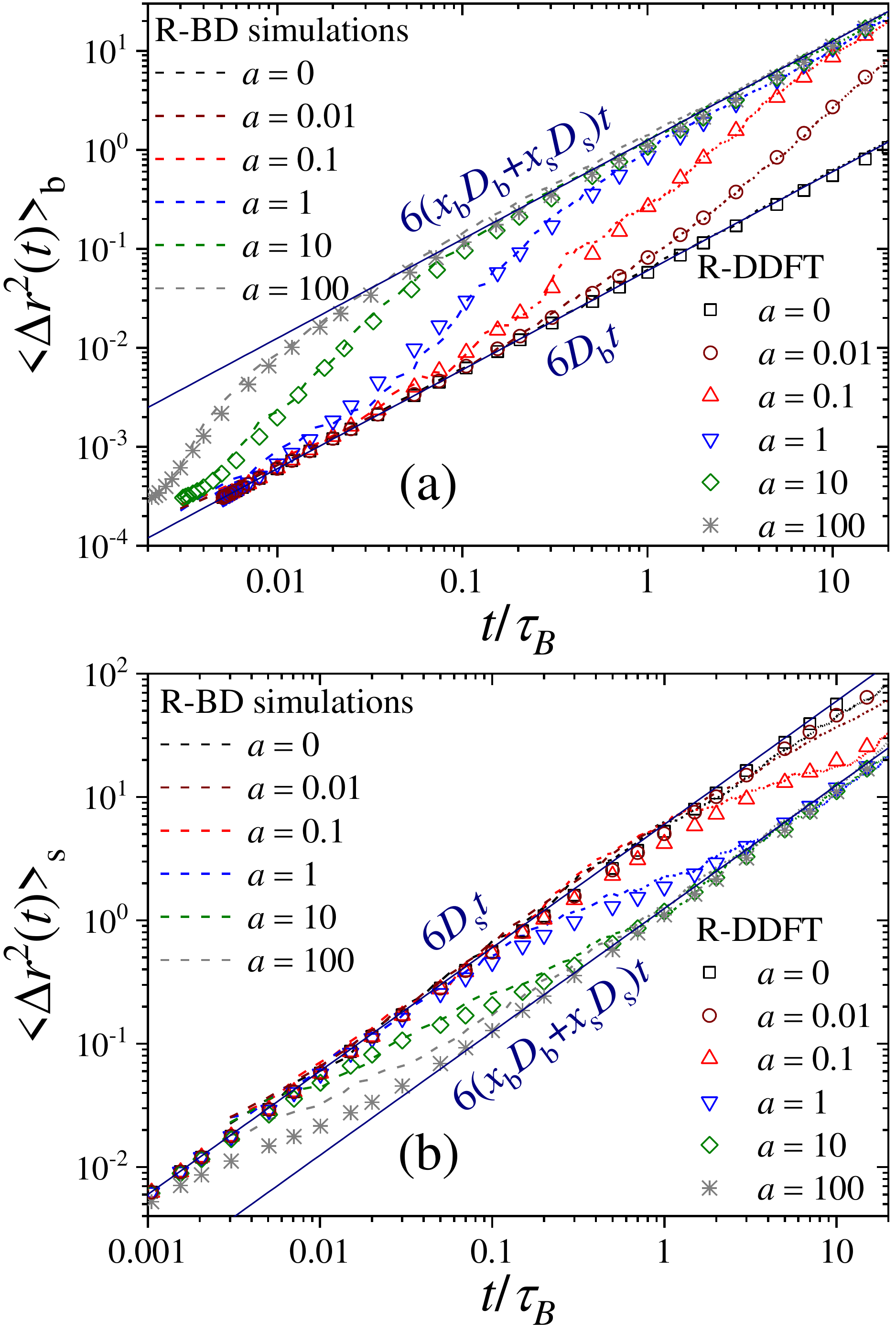}
	\caption{Comparison of the MSD obtained from R-BD simulations (dashed lines) and R-DDFT (empty symbols) for colloids initially in (a) the $\textmd{b}$-state and (b) the $\textmd{s}$-state. Calculations performed for system S2.}
	\label{fig:MSD_sim_RDDFT}
\end{figure}
\begin{figure}[ht!]
	\centering
	\includegraphics[width=0.9\linewidth]{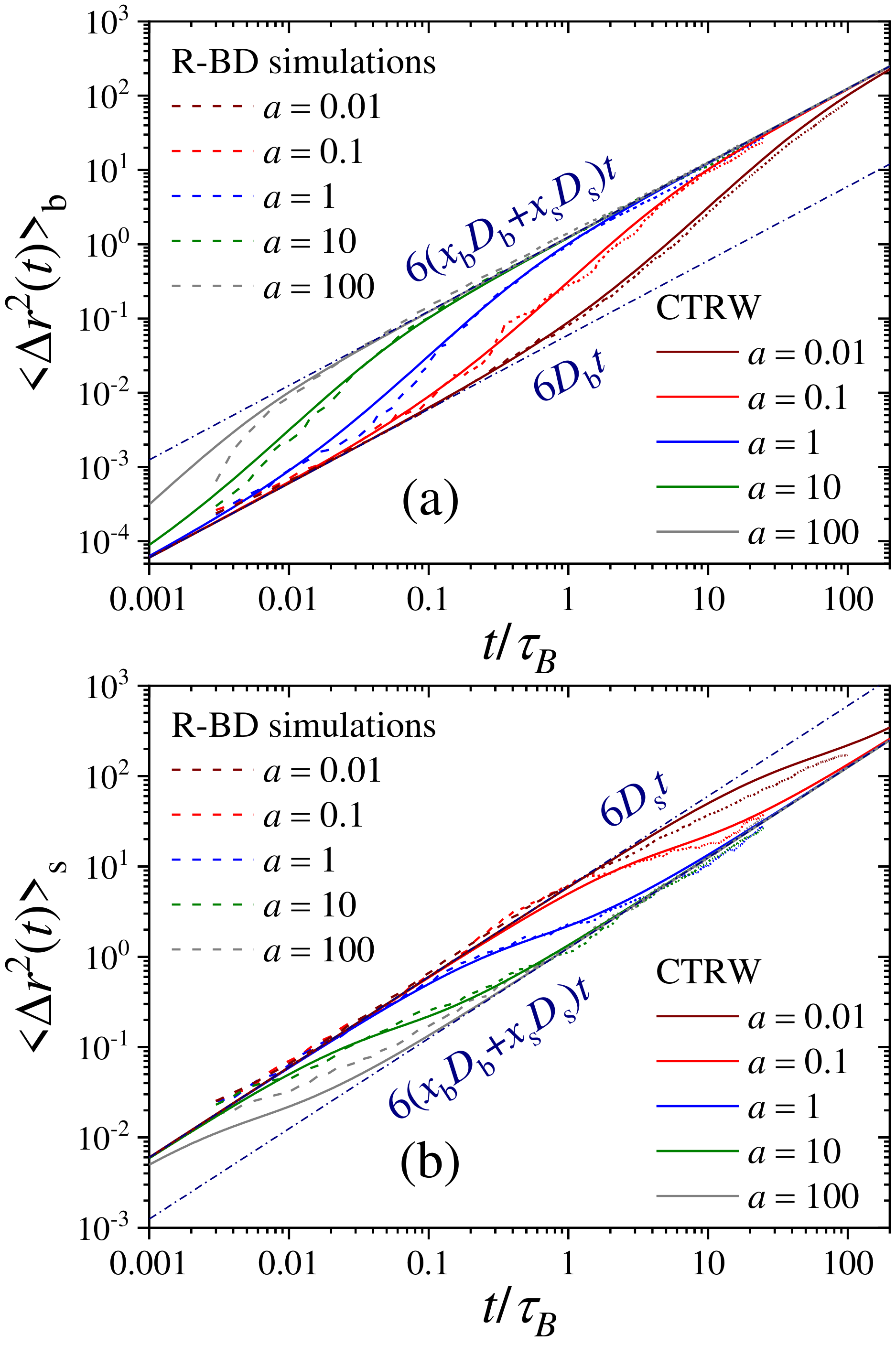}
	\caption{Comparison of the MSD obtained from R-BD simulations (dashed lines) and the phenomenological continuous-time random walk theory (solid lines) for colloids initially in (a) the $\textmd{b}$-state and (b) the $\textmd{s}$-state. Calculations performed for system S2.}
	\label{fig:MSD_sim_CTRW}
\end{figure}
In order to quantify the heterogeneous dynamics shown by the active switching particles, we now study the MSD of colloids in a particular internal state ($\textmd{b}$ or $\textmd{s}$) for different activities from $a=0$ to $a=100$. Fig.~\ref{fig:MSD_sim_RDDFT}(a) shows $\langle \Delta r^2(t) \rangle_\textmd{b}$ for a test particle initially in the $\textmd{b}$-state (system S2), obtained through Eq.~\ref{MSDb}. Empty symbols correspond to R-DDFT predictions while dashed lines are results from R-BD simulations. For $a=0$, the big test particle remains in the $\textmd{b}$-state all the time, so the MSD corresponds to a standard diffusive motion of a Brownian colloid with diffusion coefficient $D_\textmd{b}$, i.e., $\langle \Delta r^2(t) \rangle_\textmd{b}=6D_\textmd{b}t$. However, the situation changes as soon as particle switching is activated. For $a>0$, the big particle preserves its identity for times well below the switching timescale, $t \ll \tau_B/a$. Subsequently, for intermediate times $t \sim \tau_B/a$, switching events to the small (s) state significantly accelerate the dynamics of the (initially big) test particle, leading to a \emph{superdiffusive transient regime}. Finally, for large enough times $t \gg \tau_B/a$, after many switching events back and forth between the b and s states and viceversa, an effective diffusive regime is reached with an average diffusion constant $D_{ave}=x_\textmd{b}D_\textmd{b}+x_\textmd{s}D_\textmd{s}$,
\begin{equation}
\langle \Delta r^2(t) \rangle_\textmd{b}=
	\begin{cases}
6D_\textmd{b}t \ \ \ \ \ &t \ll \tau_B/a \\
6(x_\textmd{b}D_\textmd{b}+x_\textmd{s}D_\textmd{s})t \ \ \ \ \ & t \gg \tau_B/a
\end{cases}
\end{equation}
This crossover phenomenon occurs in all active systems, where individual particles show a different effective diffusion coefficient for short and long times. The transition time from one dynamic regime to the other, $\sim \tau_B/a$, is fully controlled by the activity. Indeed, for small activity rates the transition is slow so long times are required to attain the asymptotic regime, while large activity rates (such as $a=1000$) lead to a fast crossover. 

The R-DDFT theoretical predictions show an excellent agreement with R-BD simulation data (dashed lines in Fig.~\ref{fig:MSD_sim_RDDFT}(a)), thus confirming that our adapted R-DDFT represents a trustful method to investigate the dynamical properties of non-equilibrium active switching systems. Furthermore, Fig.~\ref{fig:MSD_sim_CTRW}(a) shows a comparison of the MSD for an initially-big particle obtained in R-BD simulations for different activities, and the exact prediction obtained from the CTRW theory developed in Section 3 (see Eq.~\ref{MSD_CTRW}). As for the self-intermediate scattering functions of Fig.~\ref{fig:Fs_q6}, the agreement between simulations and the phenomenological CTRW theory is also excellent. This includes both the short- and long-time asymptotic diffusive regimes with different diffusivities, but also the transient superdiffusive dynamics for initially-big particles which are accelerated by stochastic switching to the $\textmd{s}$-state.

Similar asymptotic behaviors are observed for the MSD of a test particle that was in the $\textmd{s}$-state at $t=0$, see Figure~\ref{fig:MSD_sim_RDDFT}(b). In this case
\begin{equation}
\langle \Delta r^2(t) \rangle_\textmd{s}=
\begin{cases}
6D_\textmd{s}t \ \ \ \ \ &t \ll \tau_B/a \\
6(x_\textmd{b}D_\textmd{b}+x_\textmd{s}D_\textmd{s})t \ \ \ \ \ & t \gg \tau_B/a
\end{cases}
\end{equation}
At intermediate times, however, the MSD shows now a \emph{anomalous sub-diffusive regime} caused by the slowing down of the (initially small) test particle due to switching events to the big ($\textmd{b}$) state. Both the asymptotic behaviors and the transient anomalous sub-diffusion are well-predicted by our R-DDFT theory. Moreover, Fig.~\ref{fig:MSD_sim_CTRW}(b) shows a comparison of the MSD for an initially-small particle obtained in R-BD simulations for different activities, and the phenomenological CTRW prediction, Eq.~\ref{MSD_CTRW}, and the agreement is again excellent at all timescales, including the subdiffusive anomalous regime at intermediate times.

In this way, the overall (effective) Brownian behavior of the global MSD, see Fig.~\ref{fig:VanHove2_MSDave}, results from the superposition of clearly non-Brownian dynamics of both small and big particles, which sub- and super-diffuse at intermediate times as a result of the switching activity. This anomalous diffusion properties at intermediate times are also reflected in the non-Gaussian behavior of the self-part of the van Hove displacement distribution functions (full and partial), as shown in Fig.~\ref{fig:VanHove2_Gself_bs_full}.


\section{Conclusions}

We deduced an adapted reactive mean-field dynamical density functional theory (R-DDFT), a phenomenological continuous-time random walk (CTRW) theory, and performed Brownian dynamics (R-BD) computer simulations, to investigate the dynamical properties of an active fluid of switching Gaussian colloids in the steady-state regime. For this purpose, the R-DDFT method has been generalized using the method proposed by Archer et al.~\cite{Archer2007}, in which the self-parts and the distinct-parts of the van Hove dynamic correlation function of an active system are identified with the one-body density distributions of a mixture that evolve in time according to the R-DDFT. We show that interaction switching activity not only has an important effect on the structural properties and phase behavior of the active system~\cite{MonchoJorda2020,Bley2021}, but also induces a profound change of its dynamical properties, leading to a heterogeneous diffusion.

The time-dependent distribution of displacements of particles initially in the $\textmd{b}$ and $\textmd{s}$-state (i.e. the self-parts of the van Hove correlation functions $G^{self}_\textmd{b}(r,t)$ and $G^{self}_\textmd{s}(r,t)$, respectively) follow the well-known Gaussian distribution for $t \ll \tau_B/a$, with a diffusion coefficient given $D=D_\textmd{b}$ and $D=D_\textmd{s}$, respectively. For intermediate times, $t \sim \tau_B/a$, both distributions lose the Gaussian dependence, exhibiting a well-defined crossover that separates the behavior for short and large particle displacements. This phenomenon is entirely caused by the non-equilibrium switching activity, which induces the formation of a bimodal distribution of displacements (slow/big and fast/small particles, respectively). For long enough times, $t \gg \tau_B/a$, the large number of switching events finally leads again to a Gaussian distribution, but with an effective diffusion coefficient given by the the average of the individual ones, $D_{ave}=x_\textmd{b}D_\textmd{b}+x_\textmd{s}D_\textmd{s}$. This heterogeneous diffusion is also observed in the self-intermediate scattering functions of the solution, of direct experimental relevance. In particular, a secondary relaxation plateau emerges whose height gives a measure of the fraction of colloids in the big (i.e. slow) state, and whose relaxation timescale is directly linked with the colloidal activity. 

The transition involved in the dynamics of the individual particles is also manifested in the corresponding MSDs ($\langle \Delta r^2(t) \rangle_\textmd{b}$ and $\langle \Delta r^2(t) \rangle_\textmd{s}$ for test particles in states $\textmd{b}$ and $\textmd{s}$, respectively). Indeed, particles originally in the $\textmd{b}$-state ($\textmd{s}$-state) exhibit Brownian motion with a diffusion coefficient that shifts from $D=D_\textmd{b}$ ($D_\textmd{s}$) for short times to $D=D_{ave}$ for long times. For intermediate times, $t \sim \tau_B/a$, the MSD depicts an anomalous behavior (super-diffusive for particles in the $\textmd{b}$-state and sub-diffuse for particle in $\textmd{s}$-state) connecting both dynamic regimes. Theoretical predictions obtained with R-DDFT and CTRW for the MSD of the switching interaction particles show excellent quantitative agreement with those of our reactive Brownian dynamics computer simulations (R-BD).

We believe our model applies to biological systems such as switching bistable bacteria which use switching to control structural and dynamic heterogeneity and with that collective function~\cite{Balaban, Dubnau} or synthetic realizations in, e.g., active hydrogels.~\cite{oscillating,Heuser,DNA_hydrogel,breathing}.

Our study on switching colloids inspires many future works, e.g., open questions concern the nature of the non-equilibrium thermodynamics (heat and entropy in these system), the violation of the fluctuation-dissipation theorem, and more first-principle approaches of our rather phenomenolgical model, e.g., based on active versions of the recently introduced responsive colloids (RCs) model\cite{Lin2020,Baul2021} involving continous bimodal landscape for the particle size distribution\cite{Bimodal2021}. Such a treatment may lead to position and concentration dependent switching rates with nontrivial consequences on position-dependent structure and dynamics of the active dispersions.

\section*{Conflicts of interest}
There are no conflicts to declare.

\section*{Acknowledgements}
M.B. and J.D. acknowledge support by the state of Baden-Württemberg through bwHPCand the German Research Foundation (DFG) through grant no INST 39/963-1 FUGG (bw - ForCluster NEMO) and by the Deutsche Forschungsgemeinschaft (DFG) via grant WO 2410/2-1 within the framework of the Research Unit FOR 5099 "Reducing complexity of nonequilibrium" (project No. 431945604). P.I.H. and A.M.-J acknowledge the financial support provided by the Spanish Ministry and \emph{Agencia Estatal de Investigaci\'on} (AEI) through Grants PID2020-113681GB-I00 and FIS2017-84256-P, the Junta de Andaluc\'{\i}a and European Regional Development Fund - \emph{Consejer\'{\i}a de Conocimiento, Investigaci\'on y Universidad}, (Projects PY20-00241, A-FQM-90-UGR20, A-FQM-175-UGR18 and SOMM17/6105/UGR), and the program Visiting Scholars of the University of Granada (Project PPVS2018-08). Finally, we thank Nils G\"oth for inspiring discussions and useful comments, and the computational resources and assistance provided by PROTEUS, the supercomputing center of Institute Carlos I for Theoretical and Computational Physics at the University of Granada, Spain.



\balance


\bibliography{paper_Active_Interaction_Switching} 
\bibliographystyle{rsc} 

\end{document}